\documentclass[12pt,preprint]{aastex}

\usepackage{subfigure}

\slugcomment{To appear in the Astrophysical Journal}

\shorttitle{Infrared Photometry of Starless Dense Cores}
\shortauthors{Murphy and Myers}

\begin{document}

\title{Infrared Photometry of Starless Dense Cores}

\author{D.C. Murphy}
\affil{Carnegie Institution of Washington, 813 Santa Barbara St., Pasadena,
    CA 91101}
\email{david@ociw.edu}

\author{P.C. Myers}
\affil{Center for Astrophysics, 60 Garden St., Cambridge, MA}
\email{pmyers@cfa.edu}

\begin{abstract}
Deep $JHK_{s}$  photometry was obtained towards eight dense molecular cores 
and $J-H$ vs. $H-K_{s}$ color-color plots are presented. 
Our photometry, sensitive to the detection of a 1 M$_{\sun}$,
$ 1 X 10^{6}$ year old star through $\thickapprox$ 35 - 50 magnitudes 
of visual extinction, shows no indication of the presence of star/disk systems
based on $J-H$ vs. $H-K_{s}$ colors of detected objects. The stars detected
towards the cores are generally spatially anti-correlated with core centers
suggesting a background origin, although we cannot preclude the possibility 
that some stars detected at $H$ and
$K_{s}$ alone, or $K_{s}$ alone, are not low mass stars or 
brown dwarfs ($<$ 0.3 M$_{\sun}$) 
behind substantial amounts of visual extinction (e.g. 53 magnitudes for L183B).  
Lower limits to optical extinctions are estimated for the 
detected background stars, with high extinctions 
being encountered, in the extreme case ranging up to at least $A_{V} = 46$,
and probably higher. The extinction data are used to estimate 
cloud masses and densities
which are comparable to those determined from molecular line studies.  
Variations in cloud extinctions are consistent with a systematic 
nature to cloud density distributions and column density variations 
and extinctions are found to be consistent with submillimeter wave 
continuum studies of similar regions. The results suggest that
some cores have achieved significant column density contrasts ($\sim 30$) 
on sub-core scales ($\sim 0.05$ pc) without having formed known stars.

\end{abstract}


\section{Introduction}

The first near infrared (NIR, $\lambda = 1 - 2.5$ $\mu$m) 
observations of dark clouds
were toward the Ophiuchus dark cloud complex by \citet{gra73}, who 
interpreted their observations as revealing the presence of an embedded 
star cluster. Pioneering studies such as this were hampered by their 
inability to distinguish  between background field stars and embedded 
pre-main sequence (PMS) stars, as well as by their 
comparative lack of sensitivity 
compared with modern IR array based
observations. Despite these problems, these initial surveys were
able to detect small numbers of PMS objects in dark cloud regions, evaluate 
the reddening laws, and place lower limits on cloud dust extinctions.     
Visual extinctions for dark cloud complexes
were typically found to be $\lesssim 10$ \citep{elia78,elib78,hyl81}.
\citet{hyl81} reviewed the early work in this field 
and discussed the use of the $J-H$ vs. $H-K$ color-color plot 
as a powerful technique to discriminate PMS stars from field stars, laying the 
ground work for modern studies.
The advent of NIR imaging brought a new impetus to NIR studies of dark clouds.
Extensive surveys of cooler dust clouds, some known regions of low 
to intermediate mass
star formation, have now been performed. These surveys have studied the
embedded stellar content, probed the IMF, revealed
the presence of brown dwarfs, and probed the mass distribution of 
molecular clouds \citep{eva99, lada99, lad98, wil97, lad90}. 
Recent studies by \citet{hil98} have shown that 
in Taurus-Auriga the $J-H$ vs. $H-K$ color-color plot method has an efficiency
of 60\% in identifying stars with optical signatures of accretion disks, whereas
$H-K$ vs. $K-L$ color-color plots have an efficiency of 100\%.
Therefore, while the $J-H$ vs. $H-K$ color-color plot method probably cannot 
be taken as an
absolute discriminant against the presence of star-disk systems, it still remains
a powerful diagnostic for the presence of star-disk systems in a statistical sense.
Also, in spite of these limitations, compared with longer wavelength IR observations,
the NIR has numerous inherent practical advantages such as
easier wide field coverage and ground based telescope aperture
and accessibility with acceptably low sky/telescope background.
The resilience of NIR photometry to reddening is also 
excellent, especially in the $K_{s}$ 
passband ($\lambda = 2 - 2.35$ $\mu$m), where $A_{K} \simeq A_{V}/11$.

Most studies to date have concentrated on wide area, low 
sensitivity surveys 
of large star cloud molecular complexes or in some instances 
individual optically cataloged Bok globules, while fewer systematic surveys
of dense molecular cores in the IR have been made. We present here the
results of sensitive ($K_{s} \lesssim 17.2 - 19.8$) NIR observations 
of dense cores found in dark clouds. These cores are 
lacking in associations with optically identified 
young stellar objects or embedded IRAS sources. 

An abundance of evidence now indicates that galactic star formation occurs in
cold ($T = 10 - 20$ K) dense ($n \geq 10^{4}$ cm$^{-3}$) molecular cores either 
found as components of larger molecular cloud complexes or 
as isolated dark clouds. 
Dense cores are identified by surveying 
these clouds in radio and millimeter spectral lines that trace dense 
molecular gas such as NH$_{3}$ and C$^{18}$O \citep{mye83,ben89}. Clear 
evidence exists that these regions represent a link to the earliest stages of 
star formation. For instance, dense cores are often found in close proximity to 
groups of T-Tauri stars. More directly, some dense cores are known to have 
embedded young stellar objects (YSO's) or other indicators of embedded young 
stars. Star formation tracers associated with dense cores include 
Class 0 and I IRAS sources, HH objects, H$_{2}$O masers, several 
types of emission line stars (or other low-mass 
stars with extreme spectral features and colors), 
and bipolar gas flows \citep{bei86, jij99}.  Naturally one is led 
to conjecture that dense cores 
represent an early stage of star formation: that of a self gravitating 
mass, portions of which are collapsing or 
have recently collapsed to form stars. 

If such stars are undergoing gravitational collapse, the spectral 
signature for a line of modest optical depth ($\tau \sim1$) 
is that of an apparent 
\textit{blue-shifted 
profile} \citep{hum68}. Observations of YSO's and Class 0 IRAS 
sources (often associated with dense cores) have shown significant excess of 
blue-shifted spectral profiles indicating a significant population of sources 
with inward motions \citep{zho93,zho94,mye95,gre97,mar97}. 
Most YSO's are, however, also known to be associated 
with molecular outflows (bipolar 
flows), confusing the interpretation of spectral line data. Also, some observed
inflows may not be truly "star forming" as most of the stellar mass 
might have already been accumulated. 

The molecular cores selected for study here are a small and special 
subset of the cores
from the lists of \citet{ben89} and \citet{lee01} and are of 
special interest as candidates for
the very earliest stage of low mass star formation. Specifically, these 
objects have been selected according to the
following criteria:

1) They are free of known associations with optically visible 
PMS objects, and in fact most optically visible 
stars in their denser regions (as ascertained from Palomar Observatory 
Sky Survey Plates).

2) Most are not IRAS sources and have no other indicators of embedded 
star formation. 

3) They have, in some cases, a spectral signature in millimeter wave 
transitions consistent with cloud gas infall. Five of the cores
discussed here (L1689B, L183B, TMC1, TMC2, and L158) are classified
as either "Strong Infall Candidates" or 
"Probable Infall Candidates" based a velocity analysis of extensive
mapping in the CS(2-1), N$_{2}$H$^{+}$(1-0), and C$^{18}$O(1-0) 
lines \citep{lee01}. 

4) They are well studied at other wavelengths, are relatively nearby
($d\lesssim 450$pc), and have projected molecular core sizes well
matched to the $\thickapprox$ few arcminute scales of current NIR imaging
cameras.

The objectives of our study here will be to: 

1) Provide a further sensitive diagnostic of the embedded stellar or 
brown dwarf content of the dense cores allowing a less ambiguous
interpretation of the gas velocity structure (i.e. infall vs. outflow)
and clarifying the evolutionary state of these cores. 

2) Measure the mass and density of the cores 
independent of methods based on molecular line observations which
rely on numerous assumptions regarding molecule and grain chemistry,
excitation physics, and radiative transfer.

3) Provide information on the spatial distribution and organization 
of the core gas in terms of central concentration and clumpiness.

\section{Observations and data reduction}

The observations were carried out in two observing runs  at the 1 m Swope
Telescope at Las Campanas Observatory, Chile, and one on the 60 inch telescope at Palomar Observatory. Infrared 
cameras utilizing the Rockwell NICMOS3  256 X 256 HgCdTe array were used. 
The camera used for the 60 inch telescope observations is an Offner reimager-
based, all reflective design described in \citet{mur95}. 
The infrared camera used on the 1 m Swope telescope is also an Offner reimager 
camera, very similar in design to that 
described by \citet{mur95}, differing only in optical details primarily 
related to the fact that it is deployed on a telescope with a different 
focal ratio. The field-of-view (FOV) and plate scale of both cameras is 
nearly the same at 2.6' X 2.6' and 0.60"  pixel$^{-1}$, respectively.
Table 1 summarizes the observational parameters. 

Frames at $J, H$, and $K_{s}$ were obtained for all core positions, 
with typical integration times of
several hundred seconds per filter, per core. The core positions 
(Table 2) were 
taken from the
peaks of
NH$_{2}^{+}$ and NH$_{3}$ emission line maps, both  
tracers of dense gas and velocity structure diagnostics. 
In the cases of L1709A and L1696B (which don't have
published maps), the center of
the IR frame is estimated to be approximately 50" east and 70" northeast of
the published NH$_{3}$ positions, respectively, as these positions better 
define the overall the optical core. The final $JHK_{s}$ images are 
elongated N-S from the intrinsic 2.6' X 2.6' camera fields by 
$\approx  10 - 40"$ as the telescope was dithered N-S a few times per color so that 
sky frames could be built from the data themselves as is standard with IR 
imaging. Twilight flats were taken at sunset and sometimes sunrise, and 
standard stars taken from \citet{per98} were systematically observed throughout the
night. Darks were taken for all integration times used, usually at the 
beginning and end of the night, as well as during the night if time 
allowed (such as just after a standard star measurement). 
The seeing ranged from 0.9 - 2.0" (FWHM), depending on the 
telescope, site, airmass, and night, with typical values being close 
to 1" for the Las Campanas 
observations and 1.5" being typical for the poorer Palomar 60 
inch site.  

The data were reduced using IRAF and a standard set of scripts developed for 
this purpose at the Carnegie Observatories. All data frames were first 
linearized. Following this, all dark frames, and frames taken at the same 
position in the sky were averaged with a sigma clipping algorithm 
enabled to eliminate cosmic ray hits. Averaged dark frames were
then subtracted from all data frames and the 
resultant dark-subtracted images were flat fielded using normalized twilight 
flats. 

An iterative approach to sky subtraction was used to reduce the data.
The point of the iterative operation is to suppress the residual stellar 
artifacts that would remain on a sky frame after 
conventional direct medianing, thus improving the accuracy of stellar 
photometry. 
Starting with the \emph{dithered image frames} across each core
position, i.e. linearized dark-subtracted flattened images, sky subtracted
\emph{final mosaics} were created by the following process. A
median averaged sky frame was formed from all \emph{dithered image frames}.  
This now
star-free sky frame was further flattened with Sextractor 
to form a \emph{first order sky} \citep{ber96}. This \emph{first order
sky} was subtracted from each \emph{dithered image frame} to produce 
\emph{reduced first order dither image frames}. Stars were then identified interactively on each of these
\emph{reduced first order dither image frames} and their $(x,y)$ 
positions were logged to a disk file. The
second iteration of the mosaic creation process begins by returning to the 
\emph{dithered image frames} and using the saved $(x,y)$ stellar
positions to clean them of stars to form 
\emph{star-free dithered image frames}. This 
second pass used a script based on the IRAF 
task imedit. The \emph{star-free dithered image frames} were then
medianed to obtain a \emph{final sky} for each core at each color. 
Finally, the
\emph{final sky} was subtracted from the original \emph{dithered image frames}, the resultant images were aligned and shifted 
using a fiducial reference star and sigma-clipped averaged to form a \emph{final
mosaic} at the core position. In a few cases the \emph{final mosaic}
was further processed by Sextractor to remove residual curvature.

To measure the magnitudes of the stars, the standards calibration was applied to each final 
mosaic, and 
aperture photometry
performed by hand on individual stars using an IRAF script based on the 
IRAF \emph{phot} 
routine. The brightest stars in a clean area of the frame were generally used 
to define the PSF for aperture photometry, although in cases where bright 
stars were not present, the PSF was determined from bright stars on the frames 
of other sources observed on the same night (at the same color). Aperture corrections
to the photometry of each star were estimated individually by 
graphically superimposing on the growth curve of every star the PSF. By shifting the PSF through 
a range of allowable fits, as determined by inspection, a range of possible aperture corrections 
was determined for every star. The errors in the photometry listed in Tables 3-10 are a quadrature
combination of this full range, a statistical error due to photon noise on the mosaic, and a 
sigma determined from the set of all measurements of a standard star for that
night. For faint stars, the errors are essentially dominated by the PSF fitting error, 
and for bright stars by the repeatable accuracy of measuring standard stars (since the
bright stars all had well determined PSF's). The statistical sigmas determined 
from standards measurements are based on between 9 and 24 
independent measurements per night, with an average of 13 measurements per night. These errors 
represent an absolute and well-determined floor to the photometric errors. For the fainter 
stars, we feel our error estimates tend to be on the generous side due to the approximate 
nature of our graphical fit and the fact that we roughly doubled the theoretical true error.

\section{Results}

The data are presented in the form of tables summarizing the photometry, 
processed images at each color, and $J-H$ vs. $H-K_{s}$ color-color
diagrams.

Table 2 list the approximate center positions of the
final image mosaics given in Figures 1-4 and their value in J2000.0 
coordinates. The positional errors for the 1 m 
Swope observations ($\pm 10"$ ) are higher due to the poorer pointing 
accuracy of this telescope, 
and bearing in mind that optical reference stars in the fields are not 
available because of the generally star-free nature of the sources.

Tables 3-10 summarizes the photometry for each source. The columns give
a running number identifying each star, the $JHK_{s}$ magnitudes, 
$J-H$ and $H-K_{s}$ colors, and a reddening measure. Five sigma lower 
limits are estimated for each magnitude and color and are given in
the table columns and are summarized in Table 11. To place these
detection limits in a practical perspective, the last column Table 11
presents the visual extinction A$_{V}(lim)$ through which a 1 M$_{\sun}$,
solar abundance, $ 1 X 10^{6}$ year old star should be detectable, assuming the   
core distances of Table 2, the reddening law of \citet{kor83}, and 
M$_{J} = 2.34$ \citep{sie00}.

In some cases the limiting
magnitudes are fainter for redder colors despite the fact that
observations at redder colors in the NIR are less sensitive
due to higher background for the same integration time. This 
reversal of the normal state-of-affairs was due to the lack of
detectable stars at $J$ or $H$, meaning that dithered frames could not
be stacked together so that integration times at these bluer wavebands 
were less.
Photometry and colors for a particular star are not 
given if for some reason a reliable value could not be obtained from the 
data, such as if the star is beyond or at the edge of the mosaic (specifically
noted) or for any other reason a PSF fit could not be obtained 
(denoted by ellipsis in the table entries). The stars in Tables 3-10 are 
marked in Figures 1-4 on the image mosaic used for their identification, 
usually the $K_{s}$ mosaic. 

The last column of Tables 3-10, the reddening measure $A_{V}(K5)$, is 
an attempt to provide a relative measure
of the core reddening. Calculated here is the optical 
extinction assuming that the typical background star is K5 main sequence (MS). Following the 
reddening law of \citet{kor83}, this is taken to
be
\begin{equation}
A_{V}(K5) = 15.385 E(H-K),  
\end{equation}
where $E(H-K)$ is the color excess given by
\begin{equation}
E(H-K) = (H-K) - (H-K)_{K5}.
\end{equation}
Deriving $(H-K) = 0.16$ from Table 3 of \citet{kor83}, we obtain
\begin{equation}
A_{V}(K5) = 15.385 [(H-K) - 0.160].
\end{equation}
This calculation is 
also performed for stars that have only $J-H$ lower limits and in such 
case would be a meaningful quantity if the star was 
field type as opposed to being a young star with a luminous disk embedded 
in the cloud.  We note 
that the seemingly simplistic assumption that the typical background
star is a K5 with $H-K = 0.16$ actually introduces a maximum error of only
$\approx$ $\pm2.3$ magnitudes into the extinction estimate given by (3). This is because the
intrinsic range of $H-K$ values for dwarfs and giants for spectral types A0 
to M ranges only
from 0.0 to 0.3 magnitudes \citep{kor83}. A detailed discussion of this issue is given
by \citet{alv98} and \citet{lad94}. By way of example, these authors measured extinction free 
comparison fields for the dark clouds IC5146 and L977 yielding $H-K$ = 0.13 $\pm0.01$ 
and $H-K$ = 0.20 $\pm0.13$, respectively.   

Figures 5-12  give $J-H$ vs. $H-K_{s}$ color-color plots for each core. The 
MS and reddening vectors of \citet{kor83} are plotted on each diagram.
Following \citet{lad92} we have marked the reddened positions
of stars projected off the MS at $A_{V}$ 5, 10, 15, 20, 25, and 30 
(where applicable) with X's. \citet{lad92} provide a detailed observational
and theoretical classification of YSO's in the
$J-H$ vs. $H-K_{s}$ color-color plane. Lower
limits (or in a few cases upper limits) on colors are indicated by arrows along
the direction of the undetermined color. These are mostly lower limits on $J-H$. 
In these plots, and all calculations, we have assumed that $K_{s} = K$ 
which will
introduce a negligible error in our analysis, particularly in 
light of the other uncertainties in the photometry, reddening law, and 
intrinsic spectral nature of the detected stars \citep{per98, rub98}.  
The individual objects are discussed in detail in the following sections. 

\subsection{L1709A}

Of the cores discussed here, L1709A had the most number of stellar objects
detected in the IR: Colors or limits are presented for 39 stars in Table 3.   
Figure 1 displays the $JHK_{s}$ mosaics and identifies the stars.

This core, despite its complete absence of optically identified stars
on the digitized Palomar Observatory Sky Survey (POSS),
exhibits stars in the infrared and shows 
an excess of stars with increasing infrared  
wavelength. 
The color-color plot Figure 5 indicates that 
nearly every star with 
plottable colors has colors consistent with being a reddened field star. The 
one possible exception is IR11 which is in 
a region of the color-color plot consistent with it being a 
classical T-Tauri star (CTTS) just slightly to the right of the
reddened MS. We note that even this result is suspect since this data 
point  has 
large error bars. The reddening measure (Table 3, col. 7) indicates
that the L1709A stars range from unreddened to $A_{V}(K5) = 23.6$ with
$<A_{V}(K5)> = 12.2$.

\subsection{L1582A}

L1582A was the second most populous object in terms of the number of stars
detected. Colors or limits are presented for 17 stars (Table 4). Figure 1
displays the $JHK_{s}$ mosaics and identifies the stars.

As with L1709A, this core displays numerous visible IR stars increasing
in number with wavelength, despite the fact that only two visible
stars are detected on the POSS. The color-color 
plot Figure 6 indicates that all stars with plottable 
colors but one, IR5, have colors consistent with 
being field or reddened field stars. IR5 is in the classical T-Tauri
star region (CTTS) of the color-color plot consistent with
it being a young stellar object \citep{lad92}. We note that this 
result is somewhat in doubt since this 
star is 
near the edge of the $K_{s}$ frame, probably degrading the quality of the 
photometry. The reddening measure for L1582A (Table 4, col. 7) 
varies from A$_{v} = 1.5$ to $27.9$.

\subsection{L158}

This core is devoid of stars on the POSS, but may be associated with the
Class I IRAS source 16445-1352, located approximately 2' north of the
molecular core center position and at the projected northern edge of
the core. This IRAS source is $\approx30"$ north of the northern boundary of
our IR frame \citep{lee01, bon96}. Stars are seen in our NIR frame with 
colors or limits being provided 
for eight stars in Table 5. Figure 2 displays the $JHK_{s}$ mosaic and 
identifies the stars.

The six plottable stars on the color-color diagram Figure 7  
indicate field or reddened field stars except IR6 
which displays colors consistent with it being a CTTS. 
We note however, that IR6 has very large error bars
and is faint. Furthermore the $H-K_{s}$ error bar is nearly consistent with it
being on the reddened main sequence. The $JHK_{s}$ frames clearly show a greater
visibility of stars with increasing wavelength. The reddening measure
ranges from $A_{V}(K5) = 1.0$ to $34.3$ (Table 5, col. 7).

\subsection{TMC1}

In the literature, TMC1 is often considered associated with the Class I 
IRAS 04381+2540 although we note that this object is 6' west and 1.8'
north of the molecular core center position and well outside the boundary of
our IR observations \citep{cha98}. Several stars are detected in our 
NIR frame, with
colors or limits being given for given 11 stars in Table 6.
Figure 2 displays the $JHK_{s}$ mosaics and identifies the stars.

The nine plottable stars on the color-color diagram Figure 8 
indicate that all stars have colors consistent with 
being highly reddened field stars 
although definitive statements cannot be made regarding stars with
$J-H$ lower limits. The three stars found at the rightmost portion of the
color-color diagram with $J-H$ lower limits are either highly 
reddened field stars, or PMS stars judging from the 
colors alone. The $JHK_{s}$ images show a greater visibility of 
stars with wavelength, in fact only two stars 
had measurable $J$ magnitudes.  The reddening measures ranged from 
12.6 to a remarkable 46.2, the greatest of all 67 extinction values
reported in this paper.

\subsection{L1696B}

Colors or limits are presented for 12 stars towards the L1696B core (Table 7).
Figure 3 identifies the stars and displays the $H$ and $K_{s}$ mosaics 
(no stars were detected at $J$).

Of the six plottable stars on the color-color diagram of Figure 9, 
two are clearly
consistent with being field stars, while four are either very reddened
stars or PMS objects judging from their colors alone. As is 
typical with all the cores, the images show a greater visibility of 
stars with wavelength. The reddening measures range from 18.4 to 36.0.

\subsection{L134A}

Colors or limits are presented for three stars towards the L134A 
core (Table 8). Figure 3 identifies the stars and display the 
$H$ and $K_{s}$ mosaics (no stars were detected at $J$).

All three stars are plotted on the color-color diagram Figure 10 as points with
$J-H$ lower limits. Technically these are either PMS stars 
or  highly reddened field stars. The reddening measures for 
these stars are 18.8, 16.5, 24.4.

\subsection{TMC2}

Colors or limits are presented for seven stars (Table 9) towards the TMC2
core. Figure 4 identifies the stars and displays the $H$ and 
$K_{s}$ mosaics (no stars were detected at $J$).

Only three of the stars were plottable on the color-color diagram 
(IR3, IR6, and IR7) and these only with $J-H$ lower limits making their nature 
uncertain (Fig. 11). The reddening measures for these stars are 
33.9, 23.4, and 26.6 for IR3, IR6, and IR7, respectively.

\subsection{L183B}

Colors or limits are presented for three stars towards the L183B 
core (Table 10). Figure 4 identifies the stars and displays the $H$ and 
$K_{s}$ mosaics (no stars were detected at $J$).

Only one star (IR3) was plottable on the color-color diagram, and although 
it only has a $J-H$ lower limit, its colors are consistent with it being a 
reddened field star (Fig. 12). This core is the closest to being truly
free of any associated stars 
of the sources studied here, displaying no detections at $J$ and 
only one star in the
$H$ frame. IR3 has a reddening measure of 9.7. 

\section{Discussion}

These dense cores, all initially considered to be nearly star-free in 
the optical and
IRAS data, have in fact been shown to have numerous near infrared stellar 
detections 
projected near their centers. This result indicates the marginal utility
of relying on survey optical plates to identify star-free dust clouds.

The color-color plots of these stars indicate, in the case where 
all three magnitudes are available, that nearly all have colors 
consistent with field stars, often highly reddened ones. Three 
stars discussed above have colors \emph{possibly} consistent with
an identification as PMS
objects, but the evidence for this is very weak as these stars have, in all
cases, arguably underestimated photometric errors. Reobservation 
of these stars, 
and if necessary, infrared spectroscopy on a large telescope could clarify this
situation. There is no evidence of any substantial 
embedded stellar population in any of the sources (such as a star cluster).

There are many stars in our fields that are evident only at one or two
colors ($H$ and $K_{s}$), preventing their classification by use of 
the $J-H/H-K_{s}$ diagram. Our data do not allow us to make a
definitive judgement as to the nature of these objects, although some useful
constraints can be placed on their nature. As is evident from inspection of Figures
1-4, there is a clear tendency for \textit{all} stars to be anti-correlated
positionally with the core centers suggesting that all the detected stars
are background stars. If any of the $H$ and $K_{s}$ only, or $K_{s}$ only 
detected stars are core associated they
are necessarily very young since statistical estimates
of core lifetimes range from 0.3 - 1.6 Myr, and 
star formation and core dissipation times occur within a few crossing times  
\citep{elm00, lee99}. If the one and two band detected objects 
are embedded star/disk systems, IRAS observations 
allow us to put some constraints on their properties. Assuming complete
conversion of star/disk system photons into far infrared photons, and
integrating the IRAS completeness limits (12 $\mu$m, 0.4 Jy), 
(25 $\mu$m, 0.5 Jy), (60 $\mu$m, 0.6 Jy), and (100 $\mu$m, 1.0 Jy) numerically 
between 12 and 100 $\mu$m, we obtain stellar luminosity detection thresholds 
of 0.043, 0.084, 0.12, and 0.69 L$_{\sun}$ at 100, 140, 165, and 400 pc,
respectively. These IRAS limits, taken from the
IRAS explanatory supplement, are determined for galactic latitude
$b >$ 50\degr\ and may underestimate the true completeness limits since our cores
are at lower galactic latitutes. The above
luminosities convert approximately to stellar masses of  0.08, 0.11, 0.13, and
0.3 M$_{\sun}$ assuming a 1X$10^6$ year old star, where the 
theoretical isochrones of
D'Antona and Mazzitelli as presented in \citet{wil99}, and \citet{sie00} have 
been used. Consider, for example a core at 100 pc, with a 1X$10^6$ year 
old 0.08 M$_{\sun}$ 
star which would have an $H$ absolute magnitude 
of $\approx$ 5.5 \citep{wei00}. Taking the $H$ detection limit for L183B from 
Table 11, a distance modulus of 5, and the reddening law from \citet{kor83},
indicates that such a star would necessarily suffer 53 magnitudes of
visual extinction to evade detection at $H$.
Such an extinction value, while quite high, is not precluded by our data.   

The mean extinction measure as defined above is tabulated for each case
 in Table 12. To the 
extent that this measure indicates a reasonable indication of cloud optical
extinction one can say that core extinctions range from  10 to about 30 
magnitudes with a grand average of 19.4. 
The largest mean extinction was towards TMC1 (30.3), which 
also had the most reddened
star IR7 with $A_{V}(K5) = 46.2$. Deeper and longer integrations on
larger telescopes will be necessary to penetrate to the inner regions of these
cores in the near infrared. 
These substantial extinctions clearly  
underestimate the true core extinctions, since, as is evident from the mosaic
images in Figures 1-4, the detected stars strongly avoid the central regions of
the cores. Our measured peak extinctions are a factor of $\sim 10 $ higher  
than that determined by optical star counts of dark cloud regions and
a factor of $\sim 2 - 4$ higher than that for the 
self gravitating, but non-starforming globule
B68 recently surveyed to its core in the NIR \citep{alv01, cer84}. 
Furthermore, our extinctions tend to be slightly higher than the 
wide area NIR extinction surveys of IC5146 and L977 \citep{lada99}.
Our extinctions \emph{are} generally comparable to, or less than absolute 
extinctions values found in starless prestellar cores from 
submillimeter-wave continuum surveys, consistent with the fact that 
our values are lower limits  \citep{wth99, vis01}.

The IR data allows us to estimate a lower limit to the core
densities for six cores where a clear boundary was defined on $K_{s}$ 
images by the outermost stars. The area of a polygon formed by these 
stars was calculated using core distances
taken from the literature (summarized in Table 2). The ordered star list forming the 
polygon boundary for each core is given in Table 13.  
The geometric core radius 
$\sqrt{A}$ was divided into the 
core extinction measure from Table 12, and then multiplied by the 
standard factor of 0.95X10$^{21}$ cm$^{-2}$ mag$^{-1}$ (taking $A_{V} 
= 3.1 E(B-V)$) to obtain an 
estimated lower limit on the total density of H$_{2}$ + H \citep{boh78}. 
This method also allows a rough estimate on the core mass 
lower limit given by
n$_{tot}A^{3/2}m_{H2}$, where $m_{H2}$ is the mean mass 
of an H$_{2}$ molecule in the ISM, taken to be 2.3 times the mass of 
the hydrogen atom.
The results of these calculations are given in Table 13. The densities,
masses, and radii are all typical of dense cores in dark clouds. The 
density lower limit for L1582A is in agreement with
the value of 1.6X$10^{4}$  of \citet{jij99} determined from centimeter 
wave NH$_{3}$ inversion line
data, while our density for L158 is an order of magnitude higher
than the \citet{jij99} value. 
Considering that our densities are lower limits, the L158 result
is unlikely to be due to  
geometric uncertainties, suggesting a possible breakdown of NH$_{3}$ 
molecule
as a probe of these high densities. Similar effects have been reported for
C$^{18}$O, which is found to be correlated with $A_{V}$ only up to
$A_{V} \lesssim 10$ \citep{kra99}.

Given the fact that we do not image all the way through the cores
to background stars at $K_{s}$, it is reasonable to consider 
how likely a very faint low mass T-Tauri star or brown dwarf could, if 
located near the center of the cores, evade detection. 
In Table 14 we have calculated the faintest detectable absolute 
magnitude corresponding to a star or brown dwarf 
if located at the 
center of each core. To estimate M$_{J}$, this calculation assumes the reddening law 
of \citet{kor83}, extinctions to the centers of 
half the values given in Table 12, 
cloud distances from Table 2,
and the detection limits from Table 11. Also given in Table 14 is $M_{det}$, an estimate of the
mass of a star or brown dwarf corresponding to this $M_{J}$. These masses were taken from the
evolutionary calculations given in \citet{wei00} assuming an 
upper limit of $\approx$ 1.6 X $10^{6}$ yr
for the star age based on core age estimates taken from \citet{lee99}. Considering that
the hydrogen burning limit is about 0.08 M$_{\sun}$, the results of this
simple model indicate that in all cases except for L1696B our observations can plausibly rule out
the presence of any low mass stars. In the case of L1696B, the model rules out the presence of all
but the very faintest low mass stars just above the hydrogen burning limit. In the case of L1709A
and L183B, brown dwarfs fainter than 0.01 M$_{\sun}$ should be detectable, corresponding
to objects of only 10 Jupiter masses. Our results are consistent with the
findings of the recent VLA 3.6 cm survey of \citet{har02} which
failed to detect embedded sources in four of our cores (TMC1, TMC2, L158, and L183) and placed upper limits of $\sim 0.1$ L$_{\sun}$(d/140 pc)$^{2}$ on the 
luminosities of embedded protostellar objects. We emphasize that our model results
are somewhat tentative and will require deeper NIR observations as well as follow-up $L$ band photometry to confirm.

The extinction estimates in Table 12 may
in some cases underestimate the true core extinctions and that it is possible that a star or 
brown dwarf could "hide" on the far side of a core. For instance, 
\citet{wth99} show that within a 13 arcsec beam, peak H$_{2}$ column
densities implying extinctions ranging from $A_{V} = 61$ to 
$A_{V} = 232$ are
found in prestellar dense cores similar to the sample studied here.
On the other hand, the chance that an associated star could avoid detection
is reduced if the associated star has an age in the middle of the likely
range 0.3-1.6 Myr, discussed above, instead of the upper limit on age,
1.6 Myr, we adopted earlier.
If the recent conjecture of \cite{elm00}
is correct, that star formation time occurs in a crossing time, late-type
deeply embedded stars would probably be $\approx$ 10$^{5}$ yr in age
otherwise core dissipation should have followed. Although the
evolutionary calculations given in \citet{wei00} cut on at 1 X 10$^{6}$ yr,
extrapolating from their Figure 5 (which plots $M_{J}$ vs. age) indicates
that for ages $<$ 1X10$^{6}$ yr significantly lower mass limits than 
given in Table 14 for L1709A, L1582A, TMC1, L134A, and L183B are indicated.
This is a result of the expected sharply rising intrinsic luminosity of these
objects with younger age.

The estimated individual stellar extinctions in Tables 3-10 
show considerable scatter, and generally
speaking our data are too few to make detailed statements concerning the 
nature of the clumpiness of the extincting dust. This is natural of course
as these sources were selected to be in star-free fields to begin with. 
For similar reasons it is not possible to make an estimate of cloud
density profiles. However it is clear from Tables 3-10 and Figures 1-4 that
column density contrasts ranging from $\sim$ 1.5 - 34 exist in
the clouds over scales of $\sim$ 0.05 pc, comparable to submillimeter
continuum results \citep{wth99, vis01}.    

In order to ascertain if such column density variations are random in 
nature or a result of under sampling a systematic density 
profile,  we can further analyze L1709A and L1582A, the 
two cores with enough stars to perhaps make statistical
arguments plausible and ask if our results are consistent with the 
study of \citet{ladb99} (and references therein). These studies consider 
$\sigma_{A_{V}}-A_{V}$ correlations, where $A_{V}$ is 
the mean optical extinction
calculated from all stars in a given projected area, and $\sigma_{A_{V}}$
is the standard deviation in $A_{V}$ for all stellar values 
in that same area. This relationship 
can be used to distinguish 
the true nature of the underlying continuous extinction from 
undersampled data. 

In an extensive study of the dark
cloud IC 5146 employing $\sim$ 2000 stars, \citet{ladb99} found a linear
($\sigma_{A_{V}}$, $A_{V}$) relationship.
This relationship has a slope dependent on spatial filter size. 
Based on Monte Carlo density models constructed for IC 5146, \citet{ladb99} 
showed that a smoothly decreasing density distribution falling off 
as $r^{-2}$ in a cylindrical cloud could explain 
the measured $\sigma_{A_{V}}-A_{V}$ relationships.
The spatial filter size most relevant to our data is the 90" value since this covers the largest
cloud fraction (IC 5146), as opposed to the same linear cloud size (our data sample is taken over 
the entire cloud). We note that at the distance of L1582A 
this 90" pixel is nearly the same linear dimension as in the \citet{ladb99} study since IC 5146 
is at about the same distance (400pc). 
For this spatial filter, these authors obtained  
\begin{equation}
\sigma_{A_{V}} \approx 1.0 + 0.41A_{V}.
\end{equation}
Here $\sigma_{A_{V}}$ is the mean standard deviation and $A_{V}$ the
mean optical extinction inside a 90"extinction map pixel.
These authors do not give the error bars or intercepts of their fit for the
90" pixel case, so the intercept has been estimated from their 
Figure 10 and the intercept and slope errors are assumed similar to their
equation 5, i.e. 0.11, and 0.01, respectively. 
A similar relation, but with a
slope of $0.40\pm0.02$ and intercept $1.93\pm0.11$, was obtained by 
\citet{alv98} for
a sample of 1628 stars spanning a major fraction of the optically 
visible extinction towards the dark cloud L977.
Applying the both the \citet{ladb99} and \citet{alv98} results to 
the $A_{V}(K5)$ values of
Table 12 predicts $\sigma_{A_{V}} \approx 6.0- 6.8$ for L1709A, and
$\sigma_{A_{V}} \approx 5.6 - 6.4$ for L1582A. We calculate
from all values in Tables 3 and 4, $\sigma_{A_{V}(K5)} = 5.9$    
and $\sigma_{A_{V}(K5)} = 9.2$ for L1709A and L1582A, respectively. 

Despite the considerable uncertainties in these comparisons, it is 
unlikely that the $\sigma_{A_{V}}$ values derived from our data are the result of extinction
due to either random clumpiness or a uniform smooth density distribution. For instance, 
if the extinction variation is due to completely random clumps in the line-of-sight, we
would expect an approximately Poisson distributed distribution yielding
$\sigma_{A_{V}(K5)} \approx [<A_{V}(K5)> + \sigma_{obs}^{2}]^{1/2}$. Taking  
$<A_{V}(K5)>$ values from Table 12 and a liberal estimation of $\sigma_{obs} = 2.3$ from section
3 above predicts $\sigma_{A_{V}(K5)} = 4.2$ and  
$\sigma_{A_{V}(K5)} = 4.1$ for L1709A and L1582A, respectively. On the other hand, if 
L1709A and L1582A were covered by a uniform veil of extinction with no density variation
one would expect $\sigma_{A_{V}(K5)} \approx \sigma_{obs}$, yielding 
$\sigma_{A_{V}(K5)} \approx 2.3$ for L1709A and L1582A, respectively. Both these scenarios 
predict $\sigma_{A_{V}(K5)}$ values below those measured, particularly in the case
of L1582A, tending to rule out these two possibilities. 

\section{Conclusions}

We have studied eight dense cores with deep $JHK_{s}$ photometry in the NIR,
five of which exhibit extended "infall asymmetry" in
the CS 2-1 millimeter wave transition \citep{lee01}. 
The main result of this study is, that based on the location of the 
detected stellar objects in the
$J-H$ vs. $H-K_{s}$ color-color plane, there is no substantial
evidence for an association of the cores with embedded star-disk 
systems. We note that
while our observations are sensitive enough to detect
a  1 M$_{\sun}$, $ 1 X 10^{6}$ year old star through $\thickapprox$ 35 - 50 
magnitudes of optical extinction, the data show that we have not fully
probed the core centers and a full census of the embedded star-disk system
content will require both deeper NIR and longer wavelength IR observations. 
We note as well that although (as shown by 
\citet{hil98}) the $J-H$ vs. $H-K_{s}$ color-color plane method 
is a less than perfect diagnostic of the 
presence of star-disk systems, it should statistically reveal the presence 
of such systems if they are within our detection limits unless they are very rare. 
The cores are found to have mean extinction lower limits 
of between $A_{V} \approx 10 - 30$
with individual positions showing minimum extinctions as high as
$A_{V} = 46$. The extinctions we measure are only lower limits and
probably underestimate the true core extinctions. 
These extinctions are significantly higher than typical
dark clouds or globules such as B68 determined either from optical or
NIR surveys, but are consistent with submillimeter continuum surveys
of starless cores. An analysis of individual cloud extinctions indicate
column density contrasts of $\sim$ 1.5 - 34 over scales of $\sim0.05$ pc.
A simple model suggests the cores may lack stellar objects down to at least the
hydrogen burning limit in all cases except L1696B. 
This model result should be regarded as tentative since it is based on 
a variety of assumptions 
regarding the true core extinction, 
star ages,  star cloud location, and the completeness of the $J-H$ vs. $H-K$ 
method as a diagnostic of embedded star-disk systems. 
Using the extinction data we are able to estimate core densities and masses
and have found them to be consistent with estimates determined from radio
molecular line observations.
A statistical analysis of the variation in extinction for L1709A and L1582A as characterized
by $\sigma_{A_{V}(K5)}$ is inconsistent with either a random or 
uniform density structure and may suggest a smoothly varying systematic density distribution. 

We thank Chang Won Lee for assistance with the observations. We thank
the staffs of Las Campanas and and Palomar observatories for their 
assistance. We thank Eric Persson for supplying supplying the data 
reduction scripts and providing invaluable
advice on the data reduction process. Phil Myers gratefully 
acknowledges support from NASA grant NAG5-6266.

\acknowledgments





\clearpage



\begin{figure}
\includegraphics[angle=90]{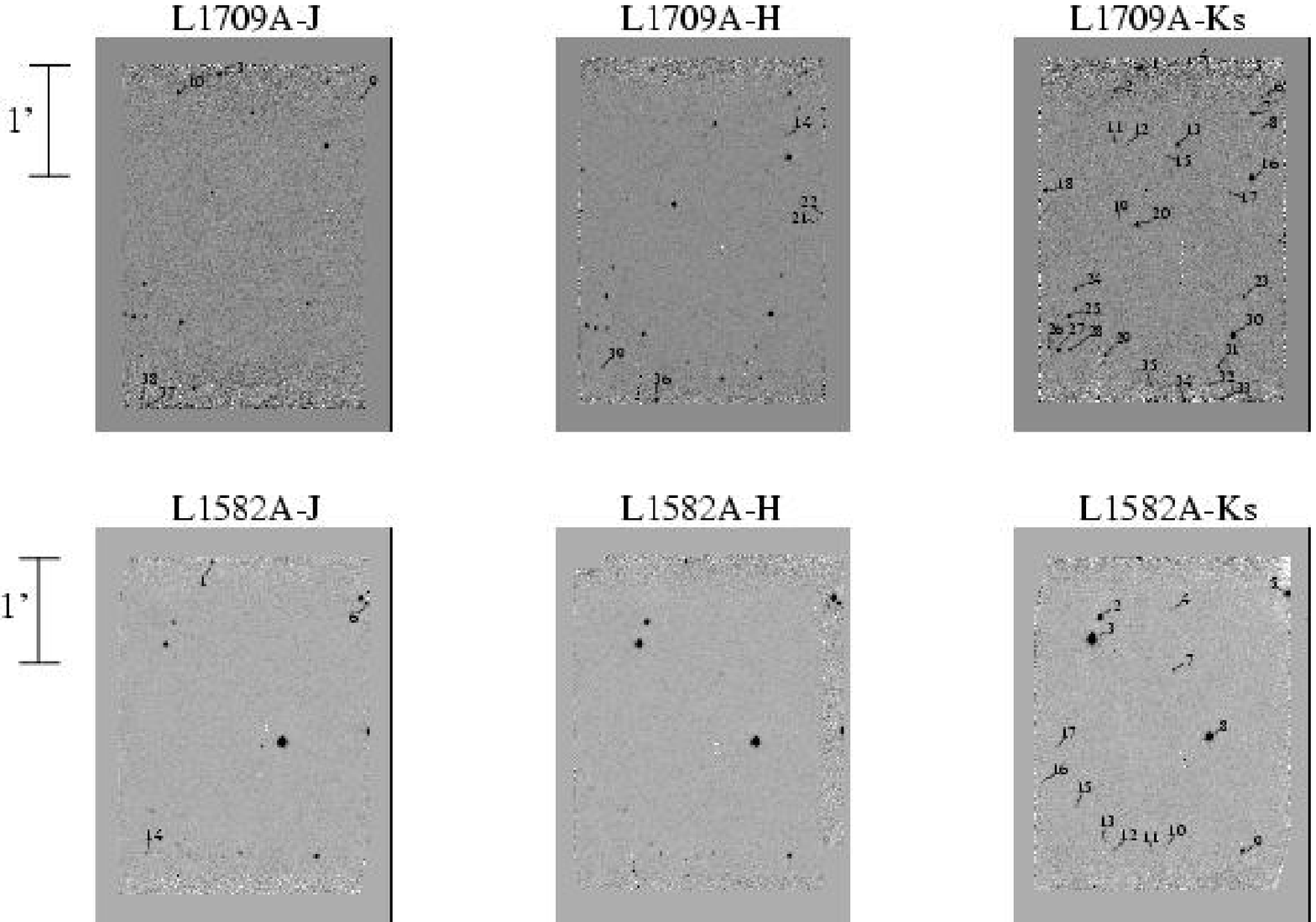}
\caption{$JHK_{s}$ mosaics of L1709A and L1582A}
\end{figure}

\begin{figure}
\includegraphics[angle=90]{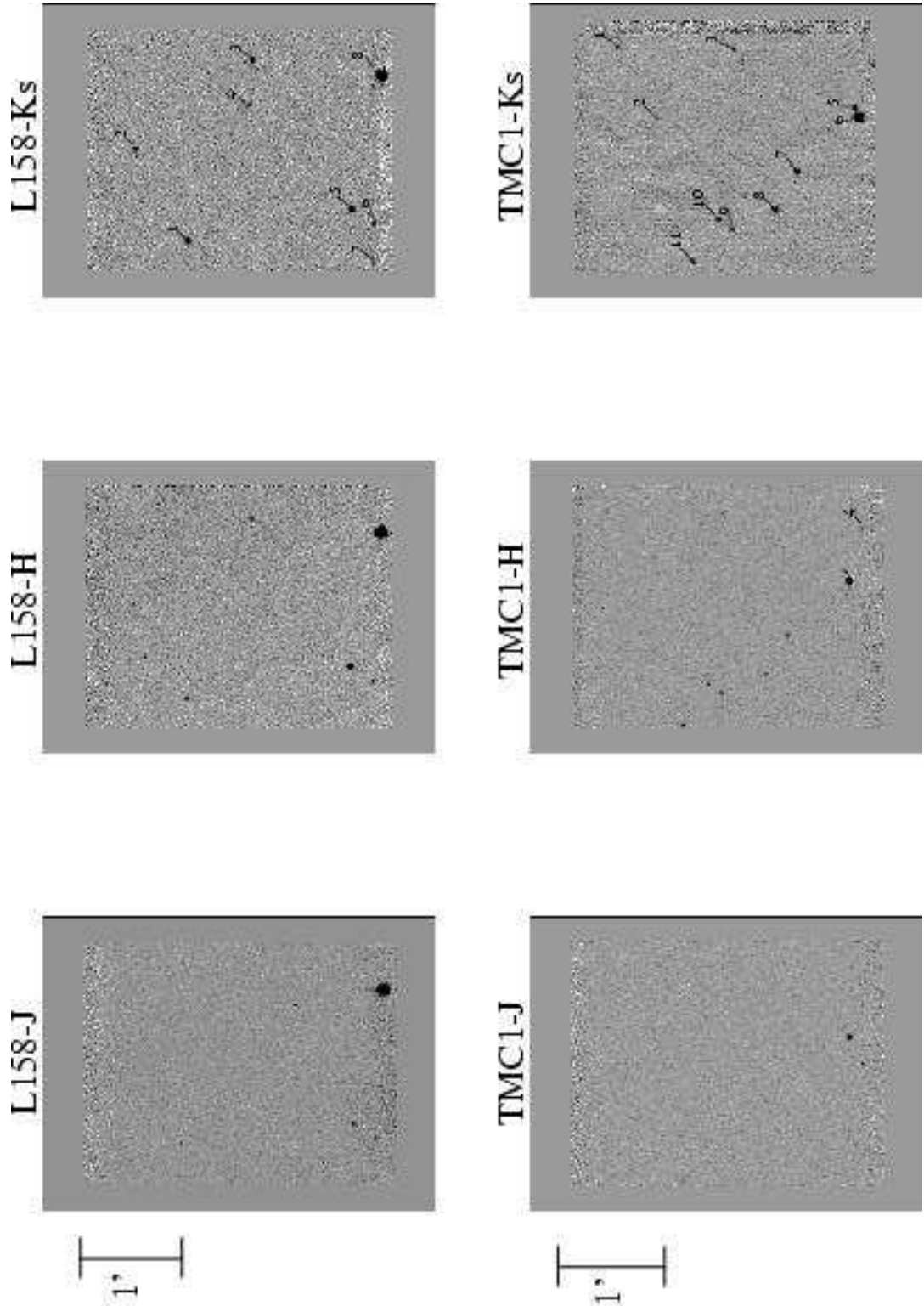}
\caption{$JHK_{s}$ mosaics of L158 and TMC1}
\end{figure}

\begin{figure}
\includegraphics[angle=90]{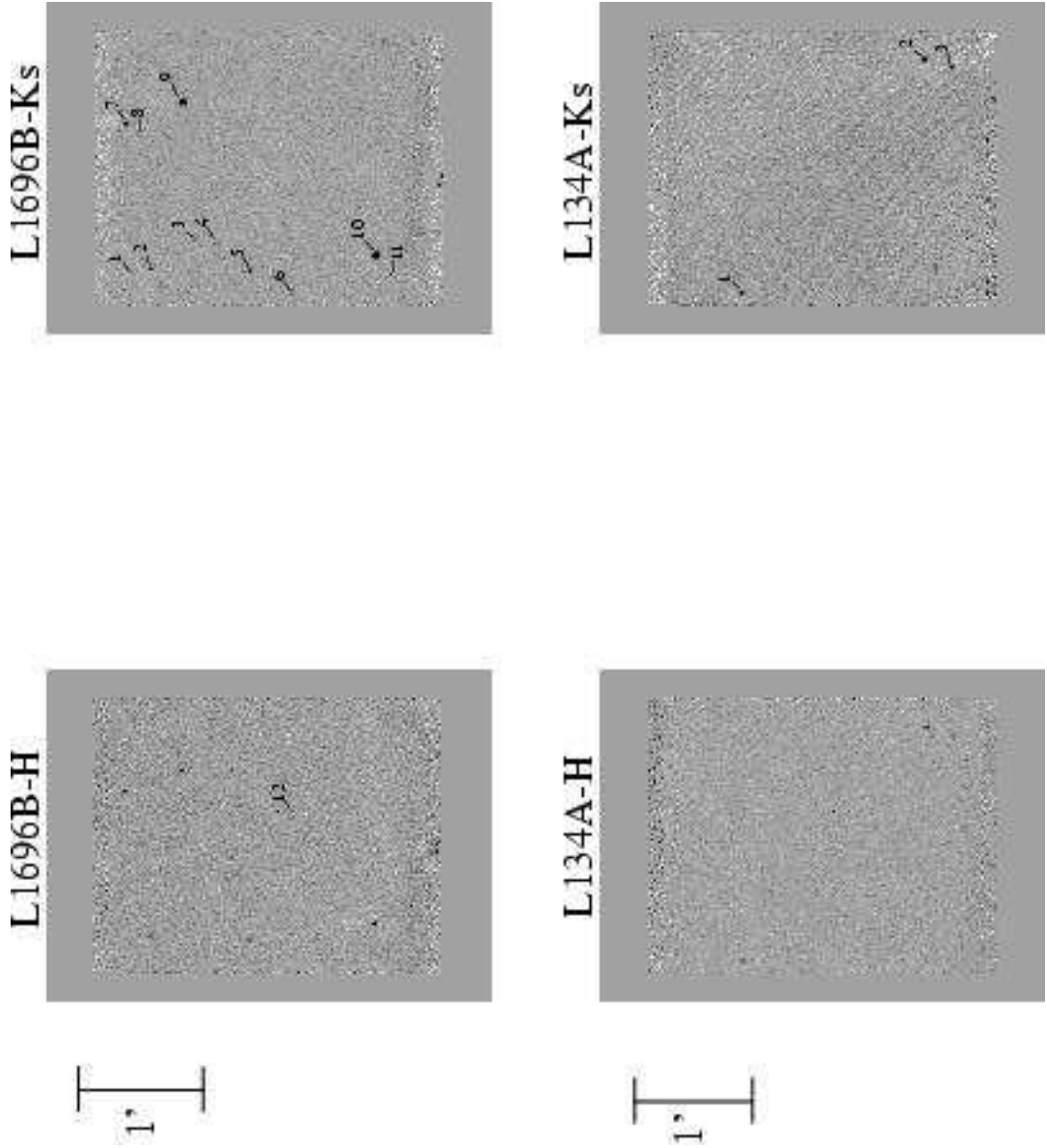}
\caption{$HK_{s}$ mosaics of L1696B and L134A}
\end{figure}

\begin{figure}
\includegraphics[angle=90]{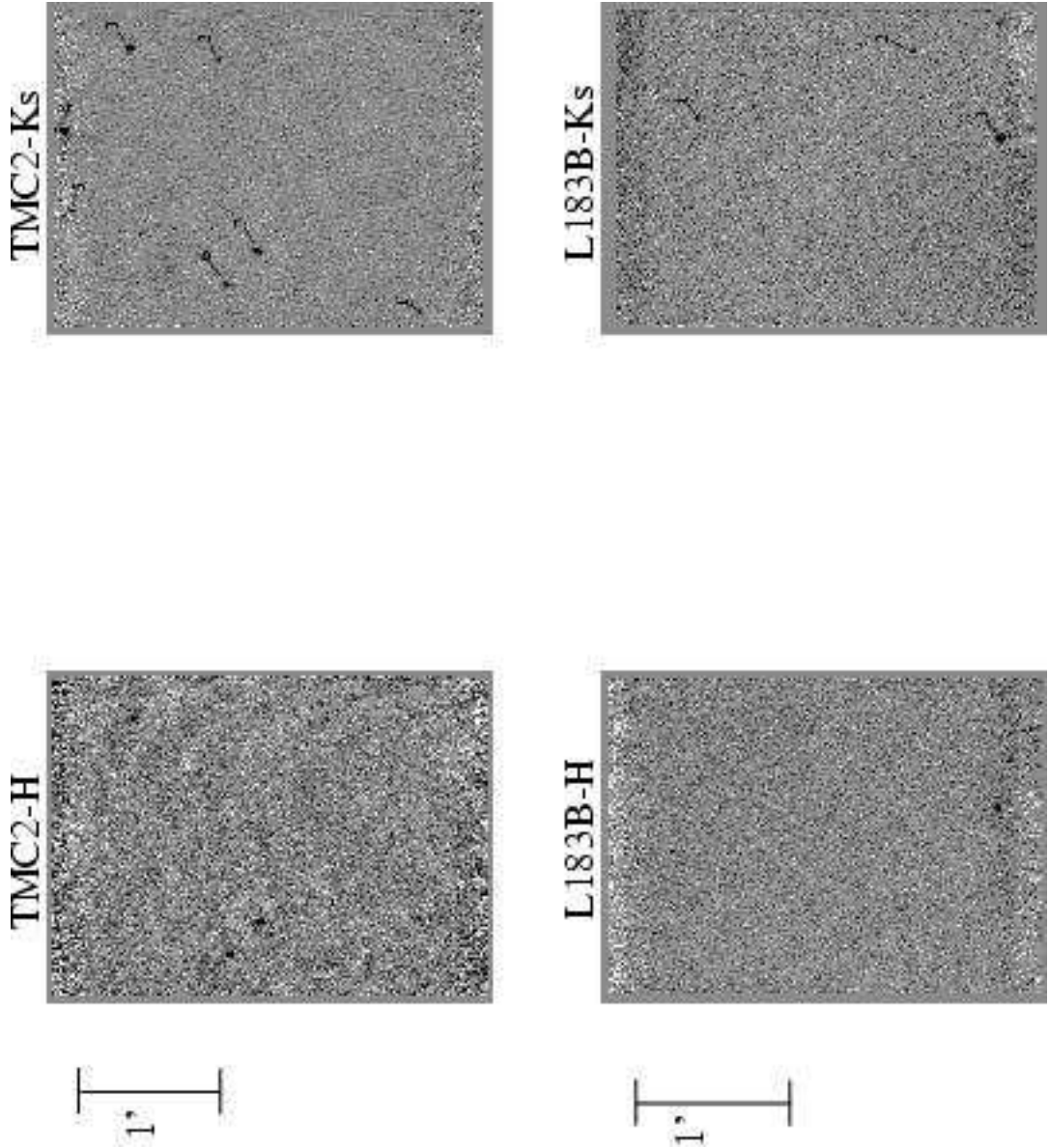}
\caption{$HK_{s}$ mosaics of TMC2 and L183B}
\end{figure}

\begin{figure}
\epsscale{1.0}
\plotone{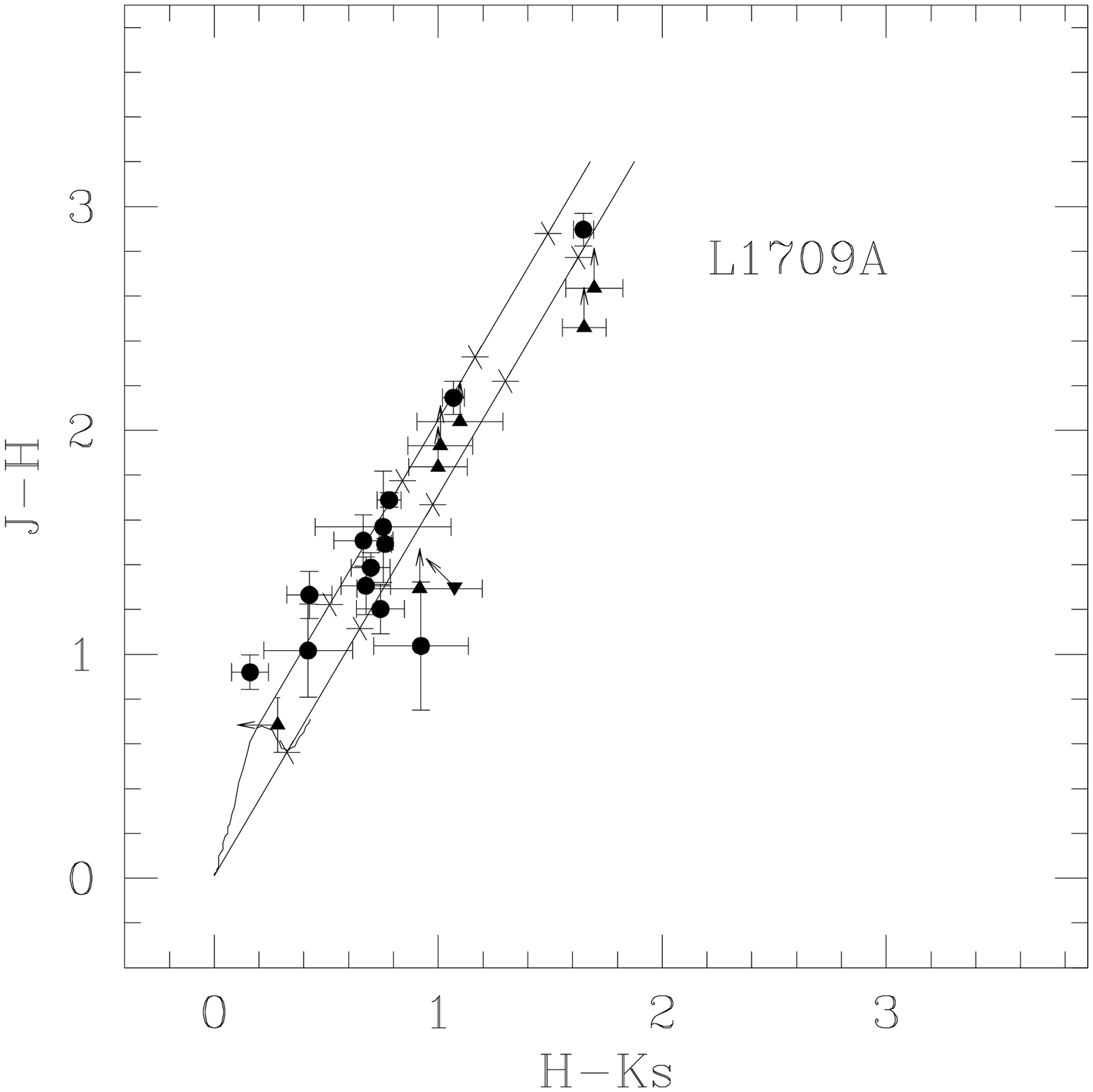}
\caption{Color-color plot for L1709A.}
\end{figure}

\begin{figure}
\epsscale{1.0}
\plotone{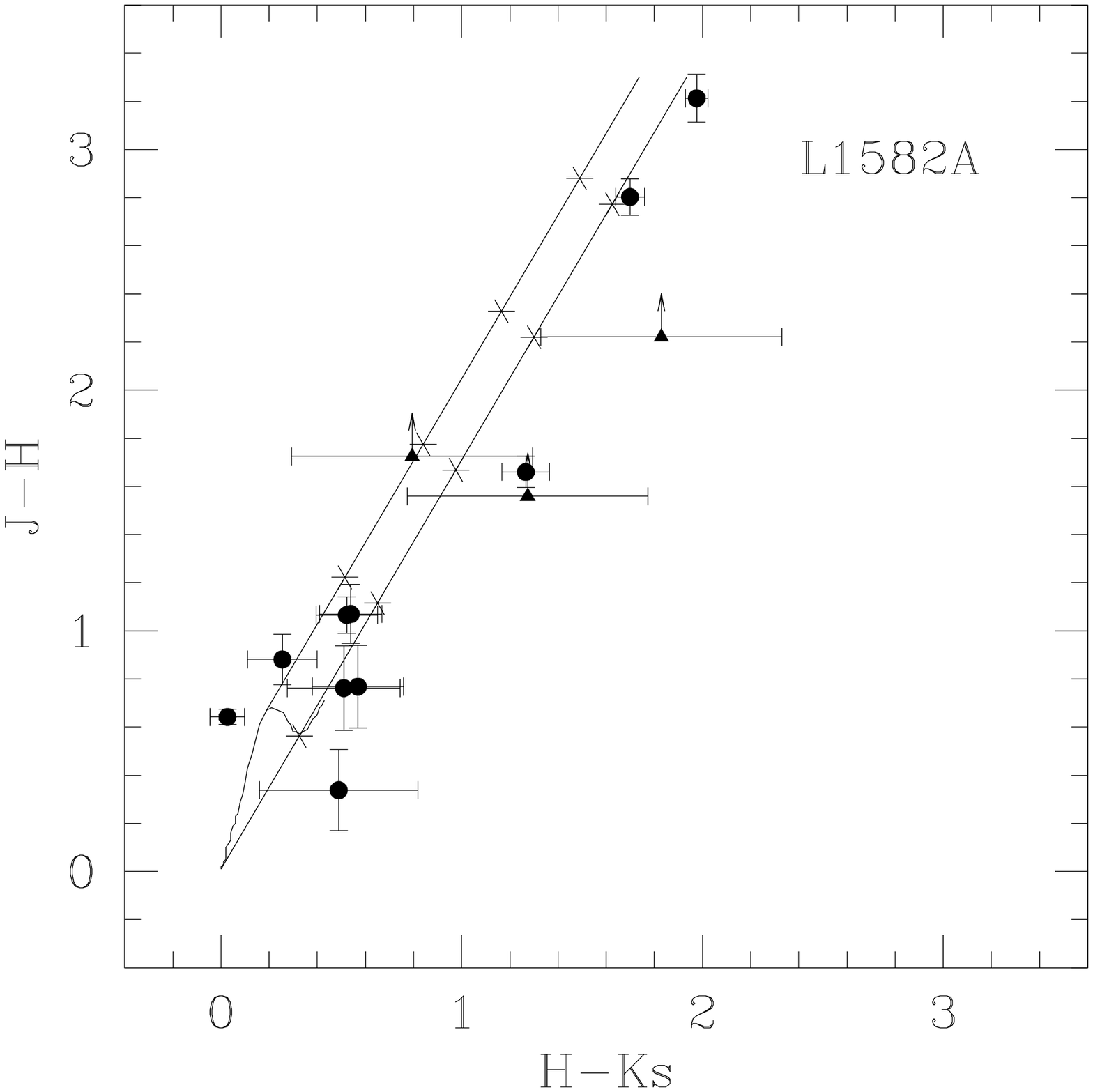}
\caption{Color-color plot for L1582A.}
\end{figure}

\begin{figure}
\epsscale{1.0}
\plotone{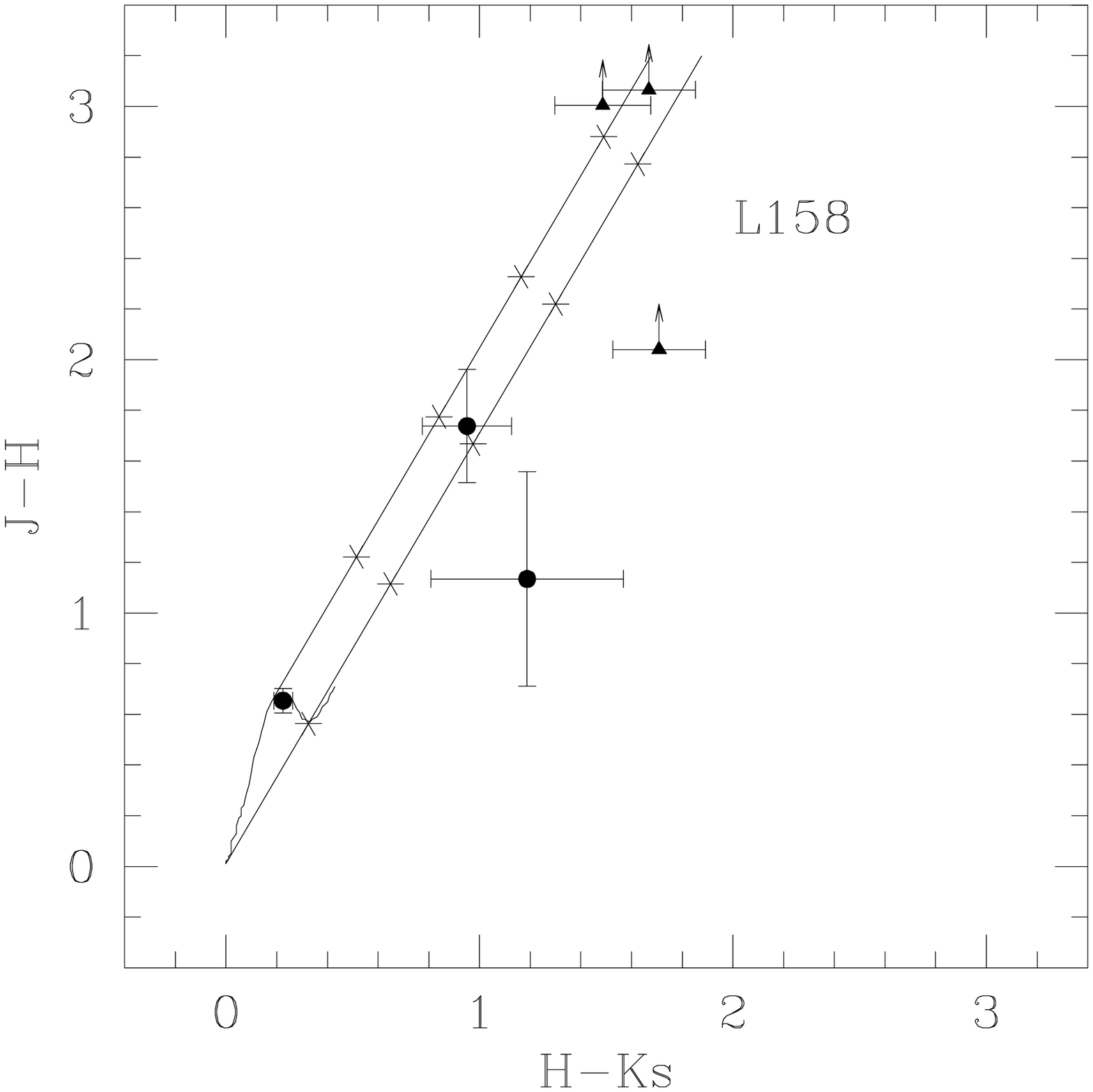}
\caption{Color-color plot for L158.}
\end{figure}

\begin{figure}
\epsscale{1.0}
\plotone{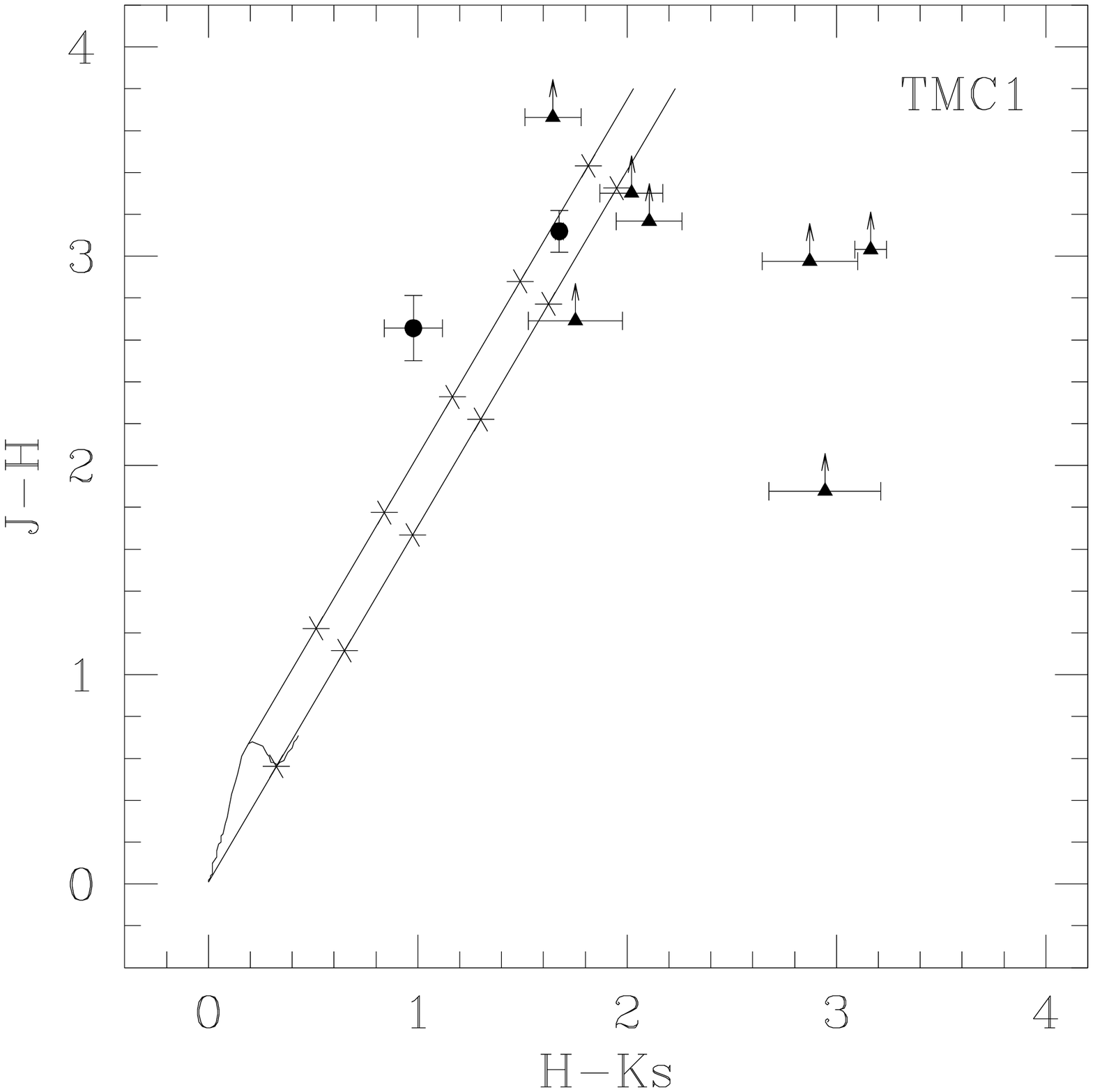}
\caption{Color-color plot for TMC1.}
\end{figure}

\begin{figure}
\epsscale{1.0}
\plotone{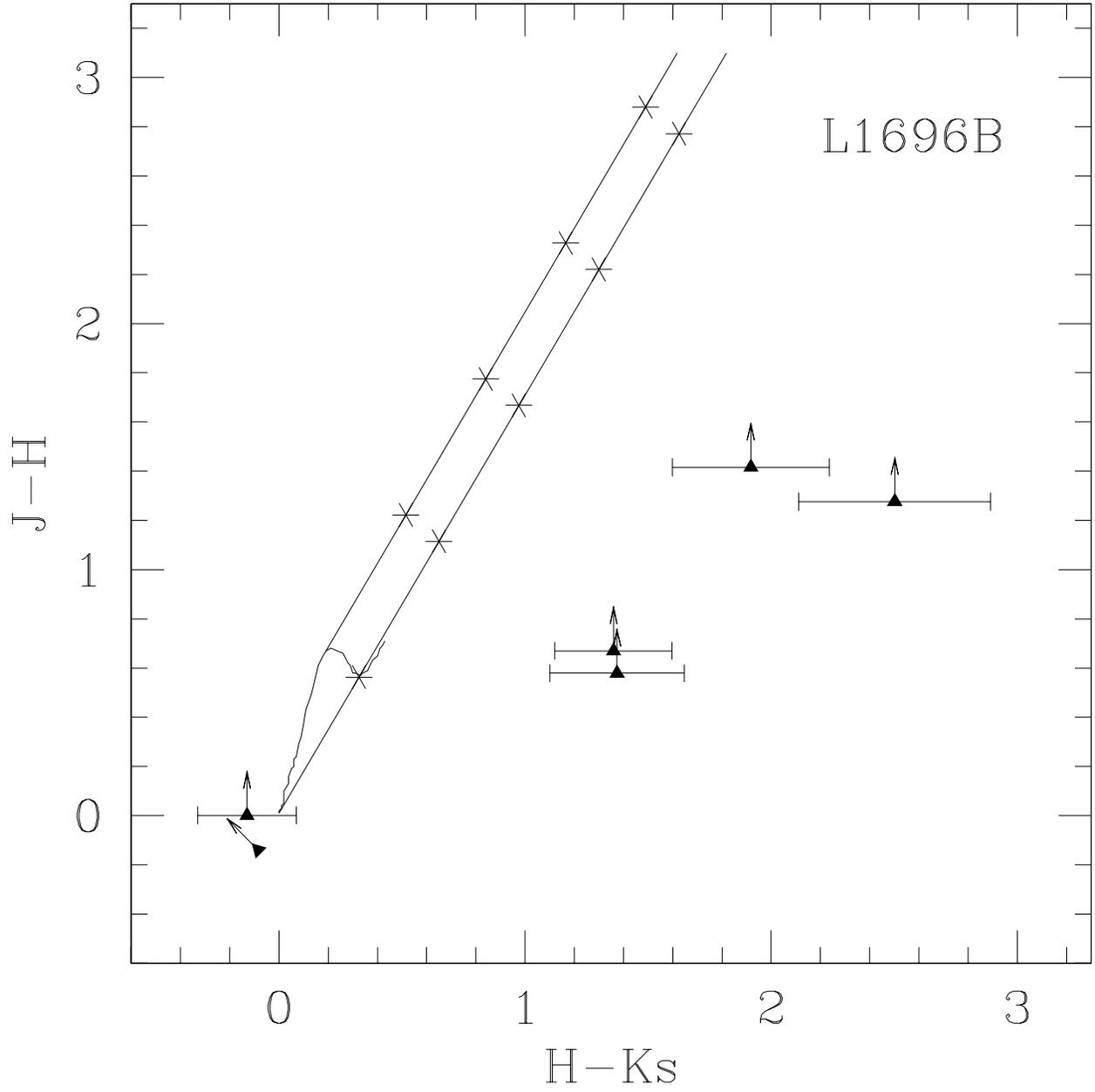}
\caption{Color-color plot for L1696B.}
\end{figure}

\begin{figure}
\epsscale{1.0}
\plotone{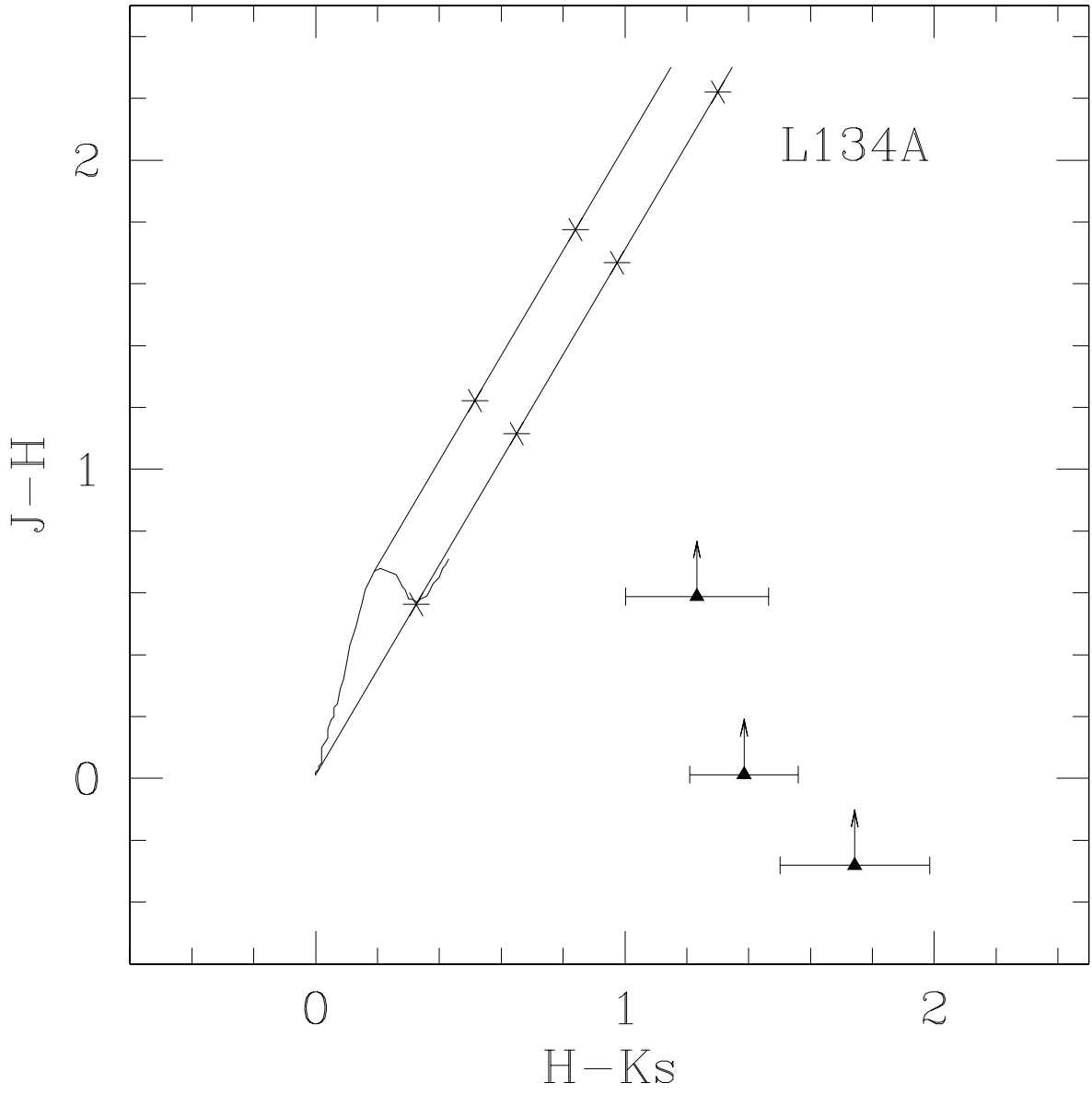}
\caption{Color-color plot for L134A.}
\end{figure}

\begin{figure}
\epsscale{1.0}
\plotone{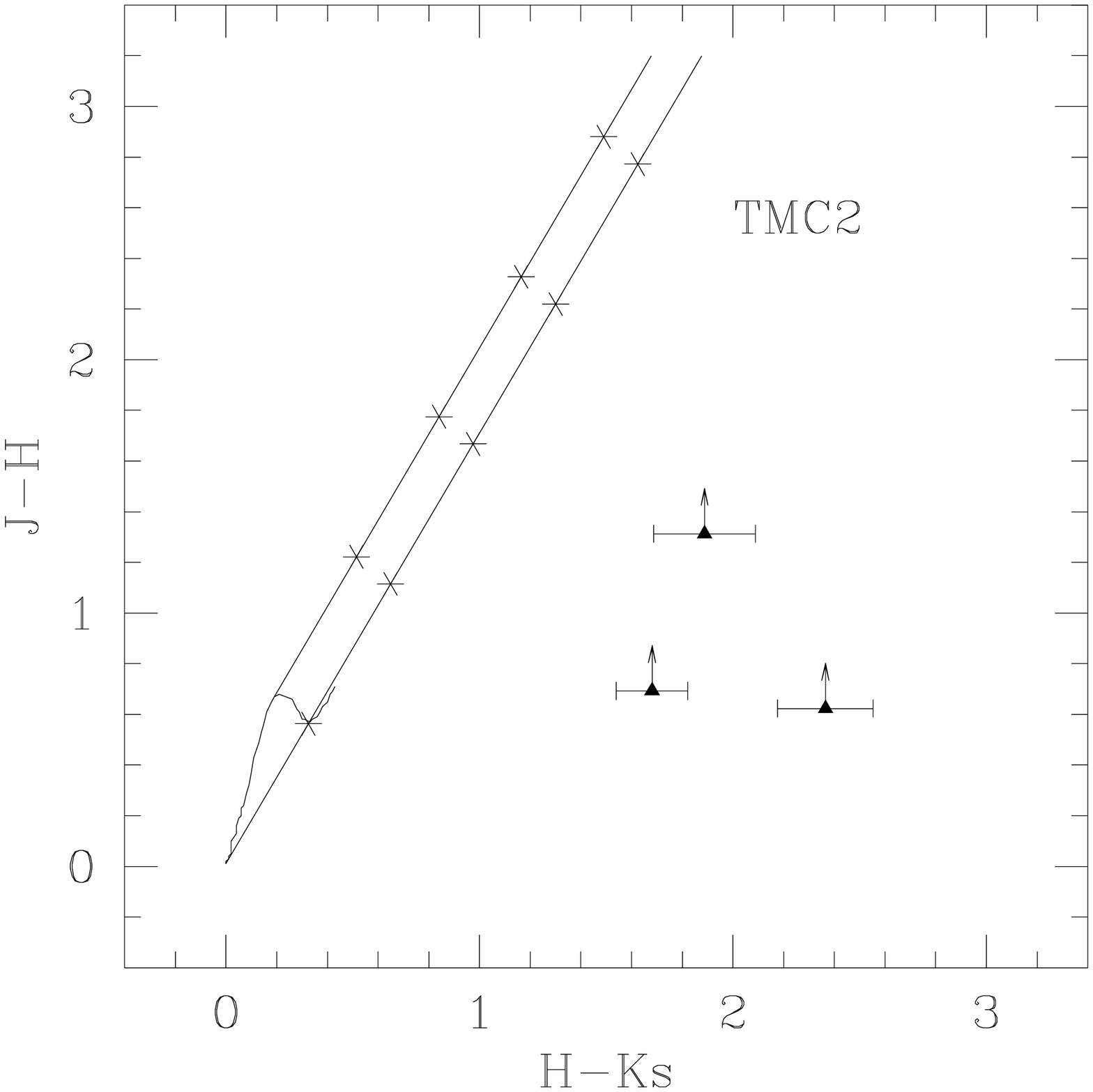}
\caption{Color-color plot for TMC2.}
\end{figure}

\begin{figure}
\epsscale{1.0}
\plotone{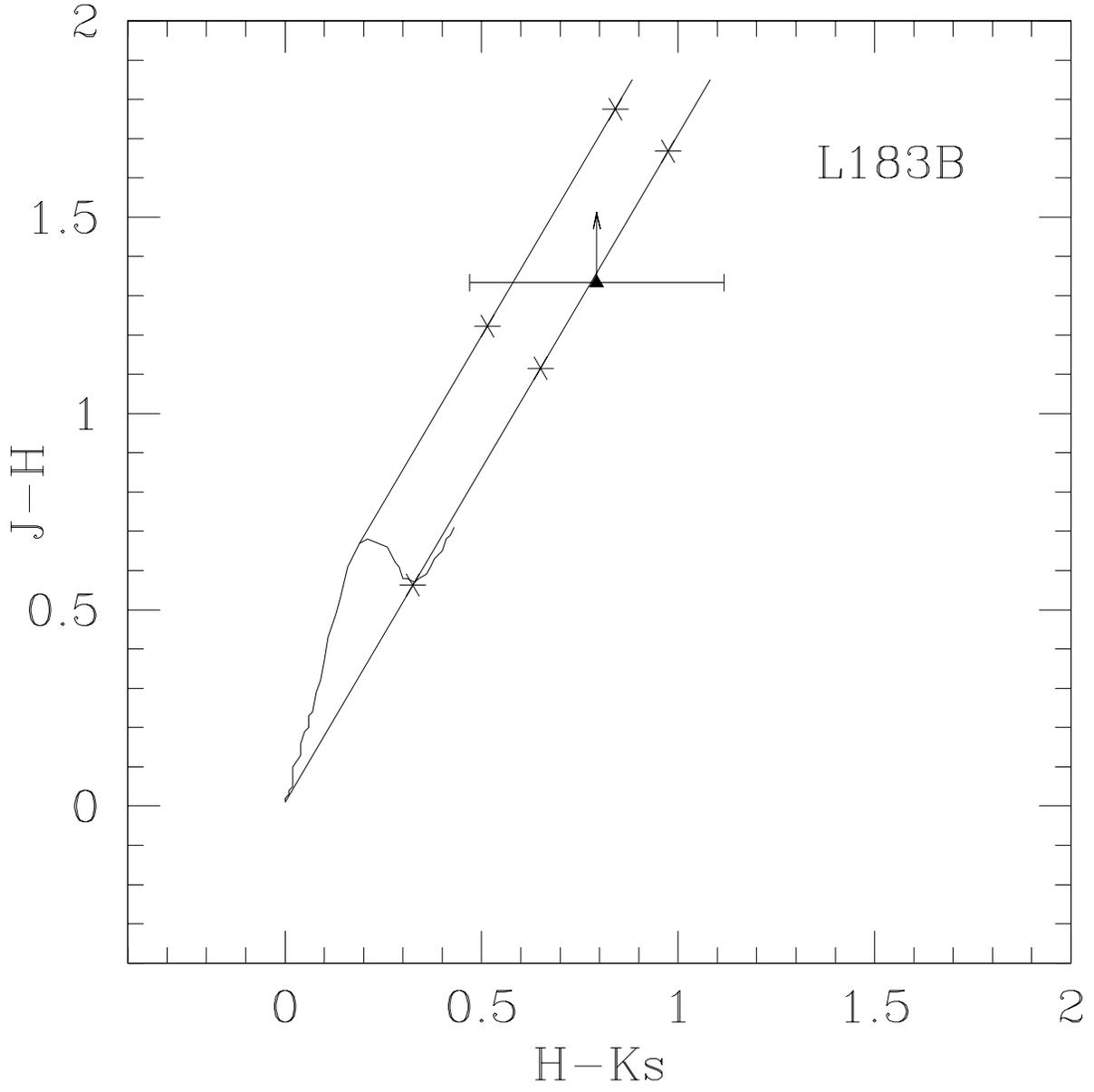}
\caption{Color-color plot for L183B.}
\end{figure}








\clearpage

\begin{deluxetable}{ccccc}
\tabletypesize{\scriptsize}
\tablecaption{Observational parameters. \label{tbl-1}}
\tablewidth{0pt}
\tablehead{
\colhead{Run} & \colhead{UT Dates}   & \colhead{Telescope}   &
\colhead{Camera Plate Scale} & \colhead{Camera FOV} 
\\
&&&(arcsec pixel$^{-1})$ &(arcmin)
}
\startdata
1 &6/17/97 - 6/23/97 &LCO 1 m &0.60  & 2.56 X 2.56\\
2 &11/6/97 - 11/8/97  &Pal 1.5 m &0.62 & 2.65 X 2.65\\
3 &3/17/98 - 3/19/98   &LCO 1 m &0.60   & 2.56 X 2.56\\
\enddata

\end{deluxetable}

\clearpage

\begin{deluxetable}{cccccccc}
\tabletypesize{\scriptsize}
\tablecaption{Starless dense core field centers and distances. \label{tbl-2}}
\tablewidth{0pt}
\tablehead{
\colhead{Core Name} &\colhead{R.A.} &\colhead{Decl.} 
&\colhead{Est. Pos. Error}  &\colhead{Position Reference} 
&\colhead{Distance}  &\colhead{Distance Reference}	&\colhead{Run}
\\
&(J2000.0) &(J2000.0) &(arcsec)  &&(pc)
}
\startdata
L1709A  &16 30 50.8   &-23 41 53.0   	&$\pm10$ 	&a	&165	&c	&3\\
L1582A  &5 32 00.3    &12 30 28.2  	&$\pm10$ 	&a	&400	&c	&3\\
L158    &16 47 23.2   &-13 59 21.1  	&$\pm10$ 	&a	&165	&c	&1\\
TMC1    &4 41 33.0    &25 44 44.0   	&$\pm5$  	&b	&140	&d	&2\\
L1696B  &16 28 59.3   &-24 20 43.0  	&$\pm10$ 	&a	&165	&c	&1\\
L134A   &15 53 36.4   &-4 35 26.0   	&$\pm10$ 	&a	&100	&e	&1\\
TMC2    &4 32 48.7    &24 25 12.0   	&$\pm5$  	&b	&140	&c	&2\\
L183B   &15 54 06.5   &-2 51 39.0   	&$\pm10$ 	&b	&100	&e	&1\\
\enddata
\tablenotetext{a}{\citet{ben89}. Also, see text.}
\tablenotetext{b}{\citet{lee01}}
\tablenotetext{c}{\citet{lee99}}
\tablenotetext{d}{\citet{jij99}}
\tablenotetext{e}{\citet{mat86}}

\end{deluxetable}

\clearpage

\begin{deluxetable}{ccccccc}
\tabletypesize{\scriptsize}
\tablecaption{Magnitudes and colors for identified L1709A stars. \label{tbl-3}}
\tablewidth{0pt}
\tablehead{
\colhead{Star} & \colhead{$J$} &\colhead{$H$} &\colhead{$K_{s}$}
 &\colhead{$J-H$} &\colhead{$H-K_{s}$} &\colhead{$A_{V}(K5)$}   
}
\startdata
1	&off frame	&off frame	&$15.037\pm0.090$  &$\ldots$  &$\ldots$ &$\ldots$ \\	
2	&off frame	& $17.752\pm0.133$	& $16.660\pm0.354$  &$\ldots$  &$1.092\pm0.378$ &$14.339$ \\	
3	& $17.692\pm0.065$	& $> 19.262$	& $> 17.868$  &$< -1.570$ &$\ldots$ &$\ldots$\\	
4	&off frame	&off frame	& $15.402\pm0.106$  &$\ldots$ &$\ldots$ &$\ldots$\\	
5	&off frame	&off frame	& $16.818\pm0.166$  &$\ldots$ &$\ldots$ &$\ldots$\\	
6	&off frame	&$16.990\pm0.127$	& $16.346\pm0.109$  &$\ldots$ &$0.644\pm0.167$ &$7.446$\\
7	&$17.626\pm0.042$	&$16.360\pm0.096$	& $15.935\pm0.032$  &$1.266\pm0.105$ &$0.425\pm0.101$ &$4.077$\\
8	&$19.172\pm0.160$	&$18.155\pm0.133$	& $17.736\pm0.147$  &$1.017\pm0.208$ &$0.419\pm0.198$ &$3.985$\\
9	&$18.741\pm0.092$	&$17.425\pm0.113$	&off frame   &$1.316\pm0.146$ &$\ldots$ &$\ldots$\\	
10	&$17.937\pm0.145$	&$> 19.262$	&$> 17.868$   &$< -1.325$ &$\ldots$ &$\ldots$\\	
11	&$19.783\pm0.244$	&$18.746\pm0.149$	&$17.823\pm0.149$   &$1.037\pm0.286$ &$0.923\pm0.211$ &$11.739$\\
12	&$20.157\pm0.235$	&$18.588\pm0.082$	&$17.834\pm0.149$   &$1.569\pm0.249$ &$0.754\pm0.304$ &$9.139$\\
13	&$17.781\pm0.047$	&$16.394\pm0.048$	&$15.695\pm0.073$   &$1.387\pm0.067$ &$0.699\pm0.087$ &$8.293$\\
14	&$> 20.243$	&$18.941\pm0.174$	&$> 17.868$   &$> 1.302$ &$< 1.073$ &$\ldots$\\	
15	&$> 20.243$	&$18.406\pm0.095$	&$17.407\pm0.090$   &$> 1.837$ &$0.999\pm0.131$ &$12.908$\\	
16	&$16.359\pm0.021$	&$14.866\pm0.016$	&$14.103\pm0.023$   &$1.493\pm0.026$ &$0.763\pm0.028$ &$9.277$\\
17	&$> 20.243$	&$18.311\pm0.111$	&$17.302\pm0.094$   &$> 1.932$ &$1.009\pm0.145$ &$13.062$\\	
18	&off frame	&$16.840\pm0.054$	&$16.021\pm0.031$   &$\ldots$ &$0.819\pm0.062$ &$10.139$\\	
19	&$> 20.243$	&$18.950\pm0.127$	&$18.033\pm0.250$   &$> 1.293$ &$0.917\pm0.280$ &$11.646$\\	
20	&$18.254\pm0.063$	&$16.108\pm0.039$	&$15.040\pm0.031$   &$2.146\pm0.074$ &$1.068\pm0.050$ &$13.970$\\
21	&$> 20.243$	&$18.337\pm0.102$	&$\ldots$   &$> 1.906$ &$\ldots$ &$\ldots$\\	
22	&$19.462\pm0.172$	&$17.918\pm0.103$	&off frame   &$1.544\pm0.200$ &$\ldots$ &$\ldots$\\	
23	&$> 20.243$	&$17.784\pm0.073$	&$16.132\pm0.064$   &$>2.459$ &$1.652\pm0.097$ &$22.954$\\	
24	&$\ldots$	&$17.358\pm0.155$	&$16.286\pm0.044$   &$\ldots$ &$1.072\pm0.161$ &$14.031$\\	
25	&$18.285\pm0.040$	&$16.777\pm0.108$	&$16.111\pm0.075$   &$1.508\pm0.115$ &$0.666\pm0.131$ &$7.785$\\
26	&$17.941\pm0.060$	&$17.021\pm0.049$	&$16.861\pm0.067$   &$0.920\pm0.077$ &$1.600\pm0.083$ &$0.000$ \\	

27	&$17.872\pm0.072$	&$16.566\pm0.106$	&$15.889\pm0.032$   &$1.306\pm0.128$ &$0.677\pm0.111$ &$7.954$ \\

28	&$18.520\pm0.063$	&$17.318\pm0.091$	&$16.576\pm0.059$   &$1.202\pm0.111$ &$0.742\pm0.108$ &$8.954$ \\

29	&$17.761\pm0.028$	&$16.072\pm0.016$	&$15.291\pm0.050$   &$1.689\pm0.032$ &$0.781\pm0.053$ &$9.554$ \\	

30	&$18.387\pm0.061$	&$15.490\pm0.041$	&$13.842\pm0.019$   &$2.897\pm0.073$ &$1.648\pm0.045$ &$22.893$ \\

31	&$> 20.243$	&$17.608\pm0.111$	&$15.911\pm0.064$   &$> 2.635$ &$1.697\pm0.128$ &$23.647$ \\	

32	&$> 20.243$	&$18.204\pm0.153$	&$17.107\pm0.116$   &$> 2.039$ &$1.097\pm0.192$ &$14.416$ \\	

33	&$> 20.243$	&$16.701\pm0.066$	&$15.406\pm0.098$   &$> 3.542$ &$1.295\pm0.118$ &$17.462$ \\	

34	&$\ldots$	&$16.972\pm0.064$	&$15.763\pm0.081$   &$\ldots$ &$1.209\pm0.103$ &$16.139$ \\	

35	&$> 20.243$	&$18.064\pm0.139$	&$16.588\pm0.085$   &$> 2.179$ &$1.476\pm0.163$ &$20.247$ \\	

36	&$16.926\pm0.019$	&$15.587\pm0.189$	&off frame   &$1.339\pm0.190$ &$\ldots$ &$\ldots$ \\	

37	&$17.438\pm0.046$	&off frame	&off frame   &$\ldots$ &$\ldots$ &$\ldots$ \\

38	&$18.700\pm0.153$	&off frame	&off frame   &$\ldots$ &$\ldots$ &$\ldots$ \\	

39	&$18.835\pm0.088$	&$18.151\pm0.085$	&$> 17.868$   &$0.684\pm0.122$ &$< 0.283$ &$\ldots$ \\	

\enddata
\end{deluxetable}

\clearpage

\begin{deluxetable}{ccccccc}
\tabletypesize{\scriptsize}
\tablecaption{Magnitudes and colors for identified L1582A stars. \label{tbl-4}}
\tablewidth{0pt}
\tablehead{
\colhead{Star} & \colhead{$J$} &\colhead{$H$} &\colhead{$K_{s}$}
 &\colhead{$J-H$} &\colhead{$H-K_{s}$} &\colhead{$A_{V}(K5)$}   
}
\startdata
1	& $16.187\pm0.97$	& $14.735\pm0.048$	&off frame  &$1.452\pm0.108$  &$\ldots$ &$\ldots$ \\	
2	& $16.837\pm0.068$	& $14.034\pm0.034$	& $12.335\pm0.050$  &$2.803\pm0.076$  &$1.699\pm0.060$ &$23.678$ \\	
3	& $15.502\pm0.094$	& $12.289\pm0.031$	& $10.313\pm0.035$  &$3.213\pm0.099$  &$1.976\pm0.047$ &$27.939$ \\	
4	& $> 19.280$	& $17.721\pm0.153$	& $16.447\pm0.113$  &$> 1.559$  &$1.274\pm0.190$ &$17.139$ \\	
5	& $14.766\pm0.028$	& $13.106\pm0.059$	& $11.840\pm0.079$  &$1.660\pm0.065$  &$1.266\pm0.099$ &$17.016$ \\	
6	& $16.162\pm0.153$	& $14.982\pm0.138$	&off frame  &$1.180\pm0.206$  &$\ldots$ &$\ldots$ \\	
7	& $> 19.280$	& $17.059\pm0.095$	& $15.230\pm0.117$  &$> 2.221$  &$1.829\pm0.151$ &$25.678$ \\	
8	& $12.070\pm0.025$	& $11.428\pm0.020$	& $11.401\pm0.048$  &$0.642\pm0.032$  &$0.027\pm0.072$ &$0.000$ \\	
9	& $15.577\pm0.047$	& $14.511\pm0.060$	& $13.988\pm0.112$  &$1.066\pm0.076$  &$0.523\pm0.127$ &$5.585$ \\	
10	& $17.429\pm0.078$	& $16.548\pm0.070$	& $16.293\pm0.127$  &$0.881\pm0.105$  &$0.255\pm0.145$ &$1.462$ \\	
11	& $17.426\pm0.124$	& $17.088\pm0.114$	& $16.599\pm0.310$  &$0.338\pm0.168$  &$0.489\pm0.330$ &$5.062$ \\	
12	& $17.195\pm0.051$	& $16.433\pm0.169$	& $15.923\pm0.162$  &$0.762\pm0.175$  &$0.510\pm0.234$ &$5.385$ \\	
13	& $17.761\pm0.156$	& $16.993\pm0.073$	& $16.424\pm0.174$  &$0.768\pm0.172$  &$0.569\pm0.189$ &$6.292$ \\	
14	& $17.831\pm0.112$	& $> 18.119$	& $> 17.170$  &$< -0.288$  &$\ldots$ &$\ldots$ \\	
15	& $17.797\pm0.113$	& $16.727\pm0.049$	& $16.188\pm0.128$  &$1.070\pm0.123$  &$0.539\pm0.137$ &$5.831$ \\	
16	&off frame	& $16.803\pm0.077$	& $16.138\pm0.069$  &$\ldots$  &$0.665\pm0.103$ &$7.769$ \\	
17	& $> 19.280$	& $17.555\pm0.102$	& $16.761\pm0.137$  &$> 1.725$  &$0.794\pm0.171$ &$9.754$ \\	
\enddata
\end{deluxetable}

\clearpage

\begin{deluxetable}{ccccccc}
\tabletypesize{\scriptsize}
\tablecaption{Magnitudes and colors for identified L158 stars. \label{tbl-5}}
\tablewidth{0pt}
\tablehead{
\colhead{Star} & \colhead{$J$} &\colhead{$H$} &\colhead{$K_{s}$}
 &\colhead{$J-H$} &\colhead{$H-K_{s}$} &\colhead{$A_{V}(K5)$}   
}
\startdata
1	& $> 19.999$	& $16.934\pm0.180$	& $15.266\pm0.046$  &$> 3.065$  &$1.668\pm0.184$ &$23.201$ \\	
2	& $> 19.999$	& $17.959\pm0.146$	& $16.250\pm0.110$  &$> 2.040$  &$1.709\pm0.183$ &$23.831$ \\	
3	& $> 19.999$	& $16.995\pm0.147$	&$15.509\pm0.119$  &$> 3.004$ &$1.486\pm0.189$ &$20.400$\\	
4	& $> 19.999$	& $> 18.577$	& $16.186\pm0.081$  &$\ldots$ &$> 2.391$ &$34.324$\\	
5	& $17.496\pm0.190$	& $15.758\pm0.118$	& $14.807\pm0.130$  &$1.738\pm0.224$ &$0.951\pm0.176$ &$12.170$\\	
6	& $18.343\pm0.253$	& $17.208\pm0.340$	& $16.020\pm0.170$  &$1.135\pm0.424$ &$1.188\pm0.380$ &$15.816$\\	
7	& $19.375\pm0.234$	& $> 18.577$	& $> 17.616$  &$< 0.798$ &$\ldots$  &$\ldots$\\	
8	& $12.738\pm0.038$	& $12.084\pm0.029$	& $11.859\pm0.023$  &$0.654\pm0.048$ &$0.225\pm0.037$  &$1.000$\\	
\enddata
\end{deluxetable}

\clearpage

\begin{deluxetable}{ccccccc}
\tabletypesize{\scriptsize}
\tablecaption{Magnitudes and colors for identified TMC1 stars. 
\label{tbl-6}}
\tablewidth{0pt}
\tablehead{
\colhead{Star} & \colhead{$J$} &\colhead{$H$} &\colhead{$K_{s}$}
 &\colhead{$J-H$} &\colhead{$H-K_{s}$} &\colhead{$A_{V}(K5)$}   
}
\startdata
1	& $> 21.396$	& $19.519\pm0.224$	& $16.574\pm0.145$  &$> 1.877$  &$2.945\pm0.267$ &$42.847$ \\	
2	& $> 21.396$	& $> 19.889$	& $17.515\pm0.078$  &$\ldots$  &$\ldots$ &$\ldots$ \\	
3	& $> 21.396$	& $18.705\pm0.143$	& $16.952\pm0.174$  &$> 2.691$  &$1.753\pm0.225$ &$24.508$ \\	
4	& $> 21.396$	& $18.746\pm0.146$	&off frame   &$> 2.650$  &$\ldots$ &$\ldots$ \\	
5	& $> 21.396$	& $17.732\pm0.093$	& $16.086\pm0.097$  &$> 3.664$  &$1.646\pm0.134$ &$22.862$ \\	
6	& $17.741\pm0.089$	& $14.622\pm0.010$	& $12.946\pm0.016$  &$3.119\pm0.099$  &$1.676\pm0.019$ &$23.324$ \\	
7	& $> 21.396$	& $18.364\pm0.073$	& $15.200\pm0.027$  &$> 3.032$  &$3.164\pm0.078$ &$46.217$ \\	
8	& $> 21.396$	& $18.420\pm0.217$	& $15.547\pm0.071$  &$> 2.976$  &$2.873\pm0.228$ &$41.740$ \\	
9	& $> 21.396$	& $18.229\pm0.093$	& $16.124\pm0.125$  &$> 3.167$  &$2.105\pm0.156$ &$29.924$ \\	
10	& $> 21.396$	& $18.095\pm0.114$	& $16.074\pm0.097$  &$> 3.301$  &$2.021\pm0.150$ &$28.631$ \\	
11	& $20.398\pm0.111$	& $17.742\pm0.110$	& $16.763\pm0.085$  &$2.656\pm0.156$  &$0.979\pm0.139$ &$12.600$ \\

\enddata
\end{deluxetable}

\clearpage

\begin{deluxetable}{ccccccc}
\tabletypesize{\scriptsize}
\tablecaption{Magnitudes and colors for identified L1696B stars. \label{tbl-7}}
\tablewidth{0pt}
\tablehead{
\colhead{Star} & \colhead{$J$} &\colhead{$H$} &\colhead{$K_{s}$}
 &\colhead{$J-H$} &\colhead{$H-K_{s}$} &\colhead{$A_{V}(K5)$}   
}
\startdata
1	& $> 18.039$	& $> 18.616$	& $17.619\pm0.294$  &$\ldots$  &$> 0.997$ &$\ldots$ \\	
2	& $> 18.039$	& $> 18.616$	& $17.131\pm0.215$  &$\ldots$  &$> 1.485$ &$\ldots$ \\	
3	& $> 18.039$	& $18.038\pm0.161$	& $18.168\pm0.120$  &$> 0.001$ &$-0.130\pm0.201$ &$\ldots$\\	
4	& $> 18.039$	& $> 18.616$	& $17.773\pm0.135$  &$\ldots$ &$> 0.843$ &$\dots$\\	
5	& $> 18.039$	& $17.369\pm0.209$	& $16.010\pm0.113$  &$> 0.670$ &$1.359\pm0.238$ &$18.447$\\	
6	& $> 18.039$	& $> 18.616$	& $17.866\pm0.101$  &$\ldots$ &$> 0.750$ &$\ldots$\\	
7	& $> 18.039$	& $17.460\pm0.249$	& $16.086\pm0.113$  &$> 0.579$ &$1.374\pm0.273$  &$18.677$\\	
8	& $> 18.039$	& $> 18.616$	& $18.200\pm0.296$  &$\ldots$ &$>0.416$  &$\ldots$\\	
9	& $> 18.039$	& $16.623\pm0.319$	& $14.706\pm0.024$  &$>1.416$ &$1.917\pm0.320$  &$27.031$\\	
10	& $> 18.039$	& $16.763\pm0.390$	& $14.261\pm0.025$  &$>1.276$ &$2.502\pm0.391$  &$36.032$\\
11	& $> 18.039$	& $> 18.616$	& $16.402\pm0.225$  &$\ldots$ &$> 2.214$  &$\ldots$\\
12	& $> 18.039$	& $18.178\pm0.167$	& $> 18.263$  &$> -0.139$ &$< -0.085$  &$\ldots$\\

\enddata
\end{deluxetable}

\clearpage

\begin{deluxetable}{ccccccc}
\tabletypesize{\scriptsize}
\tablecaption{Magnitudes and colors for identified L134A stars. 
\label{tbl-8}}
\tablewidth{0pt}
\tablehead{
\colhead{Star} & \colhead{$J$} &\colhead{$H$} &\colhead{$K_{s}$}
 &\colhead{$J-H$} &\colhead{$H-K_{s}$} &\colhead{$A_{V}(K5)$}   
}
\startdata
1	& $> 17.947$	& $17.935\pm0.163$	& $16.550\pm0.067$  &$> 0.012$  &$1.385\pm0.176$ &$18.847$ \\	
2	& $> 17.947$	& $17.359\pm0.147$	& $16.126\pm0.178$  &$> 0.588$  &$1.233\pm0.231$ &$16.508$ \\	
3	& $> 17.947$	& $18.228\pm0.122$	& $16.485\pm0.209$  &$> -0.281$ &$1.743\pm0.242$ &$24.354$\\	
\enddata
\end{deluxetable}

\clearpage

\begin{deluxetable}{ccccccc}
\tabletypesize{\scriptsize}
\tablecaption{Magnitudes and colors for identified TMC2 stars. 
The $K_{s}$ lower limit is 19.152. \label{tbl-9}}
\tablewidth{0pt}
\tablehead{
\colhead{Star} & \colhead{$J$} &\colhead{$H$} &\colhead{$K_{s}$}
 &\colhead{$J-H$} &\colhead{$H-K_{s}$} &\colhead{$A_{V}(K5)$}   
}
\startdata
1	& $> 19.329$	& $> 20.113$	& $18.336\pm0.226$  &$\ldots$  &$> 1.777$ &$\ldots$ \\	
2	& $> 19.329$	& $> 20.113$	& $17.751\pm0.069$  &$\ldots$  &$> 2.362$ &$\ldots$ \\	
3	& $> 19.329$	& $18.706\pm0.183$	& $16.341\pm0.043$  &$> 0.623$ &$2.365\pm0.188$ &$33.924$\\	
4	& $> 19.329$	& $\ldots$	& $15.576\pm0.039$  &$\ldots$ &$\ldots$ &$\ldots$\\	
5	& $> 19.329$	& $> 20.113$	& $17.192\pm0.082$  &$\ldots$ &$> 2.921$ &$\ldots$\\	
6	& $> 19.329$	& $18.636\pm0.118$	& $16.955\pm0.077$  &$> 0.693$ &$1.681\pm0.141$ &$23.401$\\	
7	& $> 19.329$	& $18.016\pm0.157$	& $16.128\pm0.124$  &$> 1.313$ &$1.888\pm0.200$ &$26.585$\\	
\enddata
\end{deluxetable}

\clearpage

\begin{deluxetable}{ccccccc}
\tabletypesize{\scriptsize}
\tablecaption{Magnitudes and colors for identified L183B stars. 
\label{tbl-10}}
\tablewidth{0pt}
\tablehead{
\colhead{Star} & \colhead{$J$} &\colhead{$H$} &\colhead{$K_{s}$}
 &\colhead{$J-H$} &\colhead{$H-K_{s}$} &\colhead{$A_{V}(K5)$}   
}
\startdata
1	& $> 18.154$	& $> 18.660$	& $17.115\pm0.207$  &$\ldots$ &$> 1.545$ &$\ldots$\\	
2	& $> 18.154$	& $> 18.660$	& $17.269\pm0.152$  &$\ldots$ &$> 1.391$ &$\ldots$\\	
3	& $> 18.154$	& $16.821\pm0.089$	& $16.028\pm0.312$  &$> 1.333$ &$0.793\pm0.324$ &$18.507$\\	
\enddata
\end{deluxetable}

\clearpage

\begin{deluxetable}{ccccc}
\tabletypesize{\scriptsize}
\tablecaption{Five sigma detection limits by filter band and extinction limits 
for each 
dense core (see text).} 
\tablewidth{0pt}
\tablehead{
\colhead{Core} & \colhead{$J$} &\colhead{$H$} &\colhead{$K_{s}$}
 &\colhead{$A_{V}(lim)$}
}
\startdata
L1709A	& $20.243$	& $19.262$	& $17.868$     & $45.3$ 	\\	
L1582A	& $19.280$	& $18.119$	& $17.170$	& $34.5$ 	\\	
L158	& $19.999$	& $18.577$      & $17.616$      & $44.5$	\\	
TMC1	& $21.396$	& $19.889$	& $19.842$	& $51.3$	\\ 	
L1696B	& $18.039$	& $18.616$	& $18.263$	& $36.8$	\\	
L134A	& $17.947$	& $20.152$      & $19.192$	& $40.7$	\\	
TMC2	& $19.329$	& $20.113$	& $19.152$	& $43.2$	\\	
L183B	& $18.154$	& $18.660$      & $19.151$	& $41.4$	\\	
\enddata
\end{deluxetable}

\clearpage

\begin{deluxetable}{cc}
\tabletypesize{\scriptsize}
\tablecaption{Mean extinction measures by core. \label{tbl-11}}
\tablewidth{0pt}
\tablehead{
\colhead{Core} & \colhead{$<A_{V}(K5)>$} 
}
\startdata
L1709A	&12.2\\
L1582A	&11.3\\
L158	&18.7\\
TMC1	&30.3\\
L1696B	&25.0\\
L134A	&19.9\\
TMC2	&28.0\\
L183B	&9.7
\enddata
\end{deluxetable}


\clearpage

\begin{deluxetable}{ccccc}
\tabletypesize{\scriptsize}
\tablecaption{Estimated core density lower limits and sizes. \label{tbl-12}}
\tablewidth{0pt}
\tablehead{
\colhead{Core}  & \colhead{n$_{tot}$} & \colhead{M} & \colhead{$\sqrt{A}$}
& \colhead{Polygon Boundary Definition}\\
&(cm$^{-3}$) &(M$_{\sun}$)		&(pc)
}
\startdata
L1709A	&5.3 10$^{4}$ & 1.3	&0.075 &15, 17, 23, 30, 31, 34, 35, 29, 25, 24, 20, 15\\
L1582A	&1.8 10$^{4}$ & 6.0  	&0.18 &3, 7, 8, 9, 10, 11, 12, 13, 15, 16, 17, 3\\
L158	&8.1 10$^{4}$ & 1.7	&0.072 &1, 2, 4, 8, 5, 1\\
TMC1	&1.6 10$^{5}$ & 2.2   &0.062 &2, 3, 5, 6, 7, 8, 9, 10, 11, 2\\
TMC2	&5.7 10$^{4}$ & 0.94   &0.065 &1, 2, 3, 4, 5, 6, 7, 1\\
L1696B  &5.1 10$^{4}$ & 0.85	&0.066 &1, 8, 7, 9, 10, 11, 6, 5, 4, 3, 2, 1\\
\enddata
\end{deluxetable}

\clearpage

\begin{deluxetable}{ccc}
\tabletypesize{\scriptsize}
\tablecaption{Results of simple model giving absolute magnitude 
detection limits to core centers \label{tbl-13}}
\tablewidth{0pt}
\tablehead{
\colhead{Core}  & \colhead{M$_{J}$} & \colhead{M$_{det}$} \\ 
& &(M$_{\sun}$)
}
\startdata
L1709A	&10.9 &$<$ 0.010	\\
L1582A	&8.4  &0.020	\\
L158	&9.0  &0.015	\\
TMC1	&7.8  &0.030	\\
L1696B  &5.4  &0.10	\\
L134A   &7.7  &0.020	\\
TMC2    &6.3  &0.070	\\
L183B   &10.6 &$<$ 0.010	\\ 
\enddata
\end{deluxetable}

\end{document}